# Universal Disorder in $Bi_2Sr_2CaCu_2O_{8+x}$

J.W. Alldredge[1], K. Fujita[2], H. Eisaki[4], S. Uchida[3], Kyle McElroy[1]

1 University of Colorado Boulder, Boulder CO

2 Cornell University, Ithaca NY

3 Department of Physics, University of Tokyo, Bunkyo-ku, Tokyo 113-0033, Japan

4 Institute of Advanced Industrial Science and Technology, Tsukuba, Ibaraki 305-8568, Japan

**Abstract**

The cuprates contain a range of nanoscale phenomena that consist of both local density of states features and spatial excitations. Many of these phenomena can only be directly observed through the use of a spectroscopic imaging scanning tunneling microscopy and their disorder can be mapped out through the fitting of a phenomenological model to the LDOS(E). The use of such a model allows a reduction of complicated LDOS($r$,E) data to key parameters whose doping and spatial dependence can be quantitatively determined. In this paper, we present a study of the nanometer scale disorder of single crystal cryogenically cleaved samples of $Bi_2Sr_2CaCu_2O_{8+x}$ whose dopings range from $p$ = 0.19 to 0.06. The phenomenological model used is the Tripartite model that has been successfully applied to the average LDOS(E) previously. The resulting energy scale maps show a structured patchwork disorder of three energy scales, which can be described by a single underlying disordered parameter. This spatial disorder structure is universal for all dopings and energy scales. It is independent of the oxygen dopant negative energy resonances and the interface between the different patches takes the form of a shortened lifetime pseudogap/superconducting gap state. The relationship between the energy scales and the spatial modulations of the dispersive QPI, static $q_1^*$ modulation and the pseudogap shows that the energy scales signatures in the LDOS(E) are tied to the onset and termination of the spatial excitations. The application of the Tripartite model allows the static $q_1^*$ modulation to have its local energy range measured and shows that its signature in the LDOS(E) is the kink, whose number of states are modulated with a wave vector of $q_1^*$. This analysis of both the LDOS($r$,E) and the spatial modulations in $q$-space show a picture of a single underlying disordered parameter that determines both the LDOS(E) structure as well as the energy ranges of the QPI, $q_1^*$ modulation and the pseudogap states. This parameter for a single patch can be defined by the Fermi surface crossing of the parent compound anti-ferromagnetic zone boundary for a model homogeneous superconductor with the same electronic properties as the patch.

**Contents:**

## 1. Introduction

An understanding of the local disorder of high temperature superconductors can provide an insight into the superconductivity mechanism. Due to the short length scale of this disorder, the only method capable of directly probing it is spectroscopic imaging scanning tunneling microscopy (SI-STM). This technique produces data sets that contain detailed information about both the spatial and energy dependence of the quasiparticle interference (QPI) of Bogoliubov quasiparticles[1–5], the checkerboard[3,6] and the pseudogap states[7–10]. It has been recently shown that the low energy empty states electronic structure (positive energies) can be classified into three energy scales using the Tripartite model[11], which is a phenomenological model explaining from both $q$-space and $r$-space SI-STM data. The Tripartite model was designed to parameterize the spatial excitations and the local density of states as a function of energy (LDOS(E)) of the underdoped low energy empty states. This model allows a quantitative description of the electronic structure and disorder as doping is changed, while matching LDOS(E) features to the three different spatial excitations. In this paper we present the low temperature real space SI-STM data interpreted using the Tripartite model for dopings ranging from 0.19 to 0.06 (OD86 K to UD20 K).

This paper is divided into seven sections, the introduction (1), background material (2), a definition of the Tripartite model (3), fitting procedure (4), fitting results (5), the relationship between the local fitting observables and the spatial modulations (6) and conclusions (7). Additional overviews of SI-STM phenomena in $Bi_2Sr_2CaCu_2O_{8+x}$ is available in several review publications[10,12]. The Tripartite model is also explained in detail and applied to mean LDOS(E) spectra and QPI gap structures in a recent publication[11].

The procedure (4) outlines the method and technology we used to fit the large data sets. The fitting results, (5), go into detail about the spatial variation in the Tripartite model parameters, how they change with doping and the fitting parameters relationship to the oxygen dopant atoms (5d). This section concludes with the single underlying disordered parameter theory (5e). In the next section, spatial modulations, $k$-space phenomena relations to local observables (6), the local fitting parameters relationship to the spatial modulations (QPI, $q_1^*$ modulation, pseudogap modulation) is explored and it is shown that the energy scales representing LDOS(E) features are related to each of the three spatial excitations. This is followed by a conclusion section (7), bringing together all the observables into one proposed phenomenological phase diagram. Complete Tripartite model fitting parameter and simultaneous topography maps for all data sets can be found in the supplementary material.

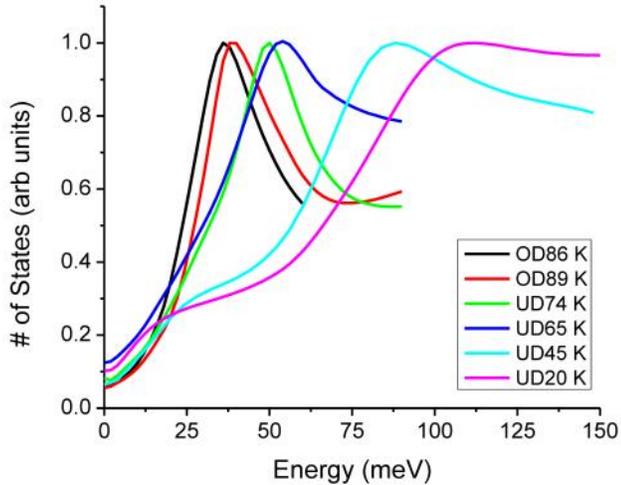

**Figure 1:** Change in the LDOS(E) with doping. These are average empty state LDOS (E) curves for the mean peak value for each doping. The curves change from a sharp peak with no kink at OD 86K to a round, suppressed peak with a large kink at UD 20 K.

## 2. Cuprate Electronic States and Phenomena Background

The low energy empty LDOS(E) of $Bi_2Sr_2CaCu_2O_{8+x}$ consists of three different regions in energy that change smoothly with changing doping. Figure 1 shows the mean LDOS(E) for six different dopings where the mean is determined using the LDOS($r$,E) peak values. At OD 86 K or p ~ 0.19 there is, on average, a d-wave $x^2-y^2$ + higher harmonic gap structure with sharp gap peaks[11,13]. As the doping is decreased, the higher harmonic contribution to the gap increases, the gap peaks becomes rounded and smaller, and there begins the growth of a plateau or kink between the low energy V-shaped LDOS(E) and the higher energy disordered LDOS peak[11,13,14]. The LDOS(E) changes continuously with doping over the range studied here and in order to reproduce the LDOS(E), a model that allows for the generation of a model LDOS(E) that has an arbitrary kink size, peak size, and ending energy of the 'V-shaped' LDOS(E) is needed. These degrees of freedom allow the model to accurately reproduce the LDOS(E) and it is necessary to provide enough degrees of freedom, without artificial constraints between them, in order to ensure that the model does not produce artificial correlations between the fitting parameters. The Tripartite model provides such degrees of freedom[11].

The low energy 'V-shaped' region of the LDOS(E) are the energies where the coherent excitations of the Bogoliubov quasiparticles from the superconducting condensate reside[10]. These excitations present themselves as the dispersive QPI, seven vectors in $q$-space that emanate from the eight ends of the superconducting band structure[1,4,15]. An example of the dispersive QPI pattern with the 7 $q$-vectors marked is shown in figure 2. The observed intensity of the $q$-vectors decrease with increasing energy and they disappear before the peak in the LDOS(E) is reached. This QPI 'termination' occurs at the energy where the kink begins[11], which is represented by $\Delta_0$ in this paper. The maximum intensity for the dispersive QPI pattern occurs at ~ 10 meV in the Z-map ($Z \equiv dI(r,+E)/dV / dI(r,-E)/dV$) for all dopings[4,5].

In the non-doped parent compound of the cuprates there is an anti-ferromagnetic ordering that is characterized by a wavelength of $(\pi/a_0, \pi/a_0)$ where $a_0$ is the atomic lattice spacing[16]. While there is no long range anti-ferromagnetic order present in the doped samples[ref], the intersection of the Fermi surface and the $(+-\pi/a_0, 0)$ to $(0, +- \pi/a_0)$ line (1st Brillouin zone parent antiferromagnetic zone boundary or PAF-zone boundary) in $k$-space coincides with the termination of the QPI[4]. The energy at which the dispersion reaches the PAF-zone boundary, also marks the onset of the kink in the LDOS(E)[11].

The LDOS(E) kink has been measured as a departure from a d-wave gap LDOS(E) in the past and has been shown to be spatial disordered[13]. However, previously it was not possible to accurately measure the location and size of the kink as it varied in space. For the mean LDOS (E) for each doping the kink grows larger in its span with decreasing doping[11] and the kinks starting energy, $\Delta_0$, decreases with energy fin the extremely underdoped samples (UD45 K, UD20 K). The kinks ending energy, $\Delta_{00}$, also increases with decreasing doping. In the Tripartite model near optimal doping the kink disappears and this is represented by a merging of the $\Delta_0$, $\Delta_{00}$ and $\Delta_1$ energy scales.

The checkerboard[3,6,17] has been described in the past as a static modulation at $q_1 \sim 2 (\pi/4a_0)$ that exists at all energy scales. It is only through the use of the Z-map that the checkerboard has been isolated and shown to begin at the QPI termination $k$-space point and energy[4,10–12]. The previous Z-map analysis revealed that the reports of the checkerboards large energy range[3,6,17] are due to setup effects inherent to the STM. Recently we have shown that the static $q_1^*$ modulation can be isolated in the raw data itself[11] and that $q_1^*$ exists at a q-vector near the dispersive $q_1$ vector, making its isolation difficult and causing confusion.

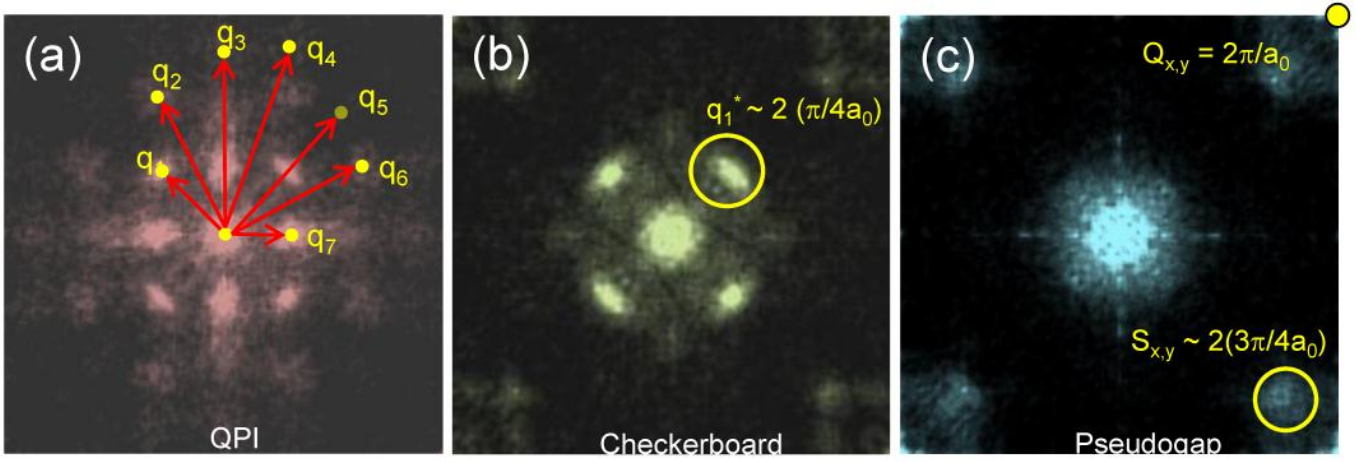

**Figure 2:** Examples of the three excitations visible in the Fourier transform of the data. These three images are taken using the Z-map, which is the ratio of the empty to the filled states in order to remove setup effects. a) shows the dispersive QPI vectors with corresponding labels. $q_5$ is not seen in the Z-map but its approximant position is marked. b) at the energy where the QPI pattern ends there is a $q_1^*$ modulation. Due to overlaps in the energy scales, $S_{x,y}$ can be seen at the $q_1^*$ modulations energies (see figure 14). c) the pseudogap pattern with its strong $S_{x,y}$ and $Q_{x,y}$ peak. Each image is undistorted to put the atomic lattice peaks in the corners at $(0, 2\pi/a_0)$. The color of each image corresponds to a particular region highlighted in figure 3.

In this paper, we discuss the $q_1^*$ static modulation that is confined to an energy range between the QPI termination ($\Delta_0$) and $\Delta_{00}$. This is what has been previously described as the checkerboard, but it is now confined to an energy range and the modulation has a well-defined signature in both the raw and Z-map data[11]. These signatures exist in energy, between the QPI and the pseudogap states. The $q_1^*$ modulation coincides with the energy range of the kink in the LDOS($r$,E)[4,5,10,11]. The intensity of the $q_1^*$ modulation decreases with increasing energy and it's -3dB point[11] occurs at the global average kink termination point, $\langle\Delta_{00}\rangle$. The $q_1^*$ wave vector is set by the intersection of the Fermi surface with the PAF-zone boundary[5,10]. An example of the $q$-space $q_1^*$ modulation pattern is shown in figure 2.

At energies above the end of the kink ($\Delta_{00}$), the high energy gap disorder begins[11,13,14]. This gap disorder takes the form of a disordered peak in the LDOS(E) at an energy of $\Delta_1$. The spatial structure of the disorder has been reported to be independent of doping and all dopings have the same normalized Gaussian distribution[13], where the Gaussians FWHM and mean value are divided by the mean value for each doping. The energy at which this gap is found has been shown to be the superconducting gap for optimally doped samples and the pseudogap for underdoped samples[18]. These two energy scales overlap or merge near optimal doping.

Coincident in energy with the $\Delta_1$ disorder are the pseudogap states that break spatial symmetry of the electronic structure[7,8]. These pseudogap states have a $q_5^*$ (or $S_{x,y}) \sim 2 (3\pi/4a_0)$ incommensurate modulation as well as commensurate $q_{Bragg}$ (or $Q_{x,y}$)= $2\pi/a_0$ that both peak in intensity at the $\langle\Delta_1\rangle$ LDOS(E) peak in underdoped samples[12]. The wavelength of $q_5^*$ is set by the intersection of the Fermi surface with the PAF-zone boundary[10]. These modulations involve broken symmetry, with the Bragg peak being a commensurate nematic signal consisting of $Q_x$ and $Q_y$ peaks at $2\pi/a_0$. $q_5^*$ is an incommensurate smectic peak consisting of $S_x$ and $S_y$. The symmetry of these excitations is not explored in this paper, but is covered in depth in previous publications[8,9]. The data sets used in this analysis lack sufficient resolution to resolve the symmetry splitting and instead, we focus on the energy dependence of the intensity in the $S_{x,y}$ state instead. An example of this pseudogap $q$-space pattern is shown in figure 2.

In the previous local fitting study[13], the LDOS($r$,E) data was fitted to a Dynes like LDOS(E) model[19] that included two separate lifetime terms. These two lifetime terms were necessary in order to match the two types of behavior of the LDOS(E) seen by high resolution, low noise SI-STM at 4 K. The first type is an increase in the number of the low energy states (energy less than approximately half the gap value) coincident with a suppression of the gap peak. These filled in gap/suppressed gap peak spectra occur around zinc impurities and other unitary scattering sites present in the sample[20]. To fit these spectra a lifetime term that is constant with energy is used and is based on weak coupling BCS unitary scattering theory[20]. The second type of the LDOS(E), representing the majority of the LDOS($r$,E) data, is a d-wave + higher harmonic gap function with a suppressed gap peak and no increase in the number of lower energy states. In order to model the suppression of the gap peak without the filling in of the gap, a linear in energy lifetime term was used. This term is modeled on Born scattering in a weakly coupled d-wave BCS superconductor[20], but the actual cause is most likely due to impurity-plus-spin fluctuations[21]. In the Tripartite model we find that both these lifetime terms are needed to match the LDOS(E).

Oxygen dopant atoms have been suggested as one of the driving forces behind the local disorder[22–24]. These dopant atoms distort the crystal structure locally and their impurity states, present at both positive and negative energies, have positive correlations with the $\Delta_1$ map. The dopant oxygen's are found in LDOS($r$,E) regions that have a larger than average high energy gap ($\Delta_1$) and smaller linear with energy lifetime term ($\alpha$)[22]. However, if the oxygen dopants are the main factor driving disorder then one would expect the $\Delta_1$ map disorder not to be universal with doping and to change significantly with changes in doping. No such changes were observed for $\Delta_1$ in a previous study[13].

There are two low energy features which are not modeled by the Tripartite model. These are the boson mode[25] seen at E ~ $\Delta_1$ + 52 meV and the low energy (< 5 meV) resonance of the zinc atoms[26]. Both of these effects are strong only in optimally doped samples and the boson mode occurs at energies higher than the $\Delta_1$ gap and therefore will not interfere with the Tripartite model fits. The zinc atom resonances only occur in a very small portion of the field of view and therefore have minimal effect on the fitting. In addition, the zinc states are well described by strong scattering models[27], which are outside the scope of the Tripartite model. The zinc atoms and vacancies location and area of effect can be seen in supplementary figures 1-9 $\Gamma_1$ maps, which provide a measure of the unitary scattering.

## 3. Tripartite Model

Due to the wide range of possible spectra combined with background noise, a model is necessary in order to quantitatively parameterize the LDOS(E). The Tripartite model is used in order to map the disorder of the local energy scales and lifetimes. This model has previously been successfully applied to mean LDOS(E) spectra and the QPI $k$-space gap structure over several dopings[11]. The Tripartite model is based off the standard superconducting spectral density of states, A($k$,E)[28] and includes both an energy independent and a linear energy dependent lifetime terms[13]. The spectral density of states used is:

$$A(\vec{k},E) = \frac{-1}{2\pi} \frac{\mathrm{Im}\Sigma(\vec{k},E)}{(E - \varepsilon(\vec{k}) - \mathrm{Re}\,\Sigma(\vec{k},E))^2 + \mathrm{Im}\,\Sigma(\vec{k},E)^2} \qquad 1$$

With the self-energy[29] calculated by

$$\Sigma(\vec{k},E) = -i\Gamma + \frac{\Delta_k^2}{E + \varepsilon(\vec{k}) + i\Gamma} \qquad 2$$

The lifetime term ($\Gamma$) is the same that was found to be necessary in the previous local fitting work[13]

$$\Gamma = \Gamma_1 + \Gamma_2(E) \qquad 3$$

$$\Gamma_2(E) = \alpha \cdot E \qquad 4$$

In past analysis's, the QPI dispersion was fit with the addition of a higher harmonic term to the d-wave gap[3,11,13,30,31]. This higher harmonic d-wave contribution takes the form of $\Delta_1*(B*\mathrm{Cos}[2*\theta]+(1-B)*\mathrm{Cos}[6*\theta])$, where B varies from 0.7-1.0. This gap contrasts with previous fitting[13] of the LDOS($r$,E), which included only a lowest harmonic d-wave gap, $\Delta_1*\mathrm{Cos}[2*\theta]$. In figure 3b, the dotted white line shows a lowest harmonic d-wave LDOS(E). In order to unify the QPI and LDOS measurements, the Tripartite model uses a higher harmonic gap term for the nodal section of the Fermi surface and this term ends at an angle, $\theta_{\mathrm{cross}}$, where the higher harmonic d-wave gap is replaced by a lowest order d-wave gap. This results in a gap structure that follows the equation:

$$\Delta[\theta] = \begin{cases} \Delta_1 \cdot \mathrm{Cos}[2\theta] & \text{for } \theta > \theta_{\mathrm{cross}} \\ \Delta_1 \cdot (B \cdot \mathrm{Cos}[2\theta] + (1-B) \cdot \mathrm{Cos}[6\theta]) & \text{for } \theta < \theta_{\mathrm{cross}} \end{cases} \qquad 5$$

This gap structure allows the generation of a spectral density of states that fits both the LDOS(E) and the QPI data using the same parameterization[11]. The region between these two gaps is represented by a non-dispersive transition. However in order to avoid double valuing the greens function this transition disperse slightly and the upper limit on the amount of dispersion is set by the resolution in measuring $q_1^*$ in the SI-STM data. This non-dispersive transition is the cause of $q_1^*$ in the Tripartite model[11] and measurements[5] of $q_1^*$ show that it has a width of ~$0.2\pi/a_0$ and is non-dispersive to ~$0.03\pi/a_0$. In figure 3a,b the **k**-space location of the peak in the Tripartite spectral density of states is shown. This maximum in **k**-space of the spectral density of states is used to fit the dispersive QPI extracted **k**-space points and an example of the result of such a fit is shown in figure 3a. The LDOS(E) is generated by integrating the spectral density of states over **k**-space and an example with a lowest harmonic d-wave gap overlaid on top of the Tripartite gap is shown in figure 3b.

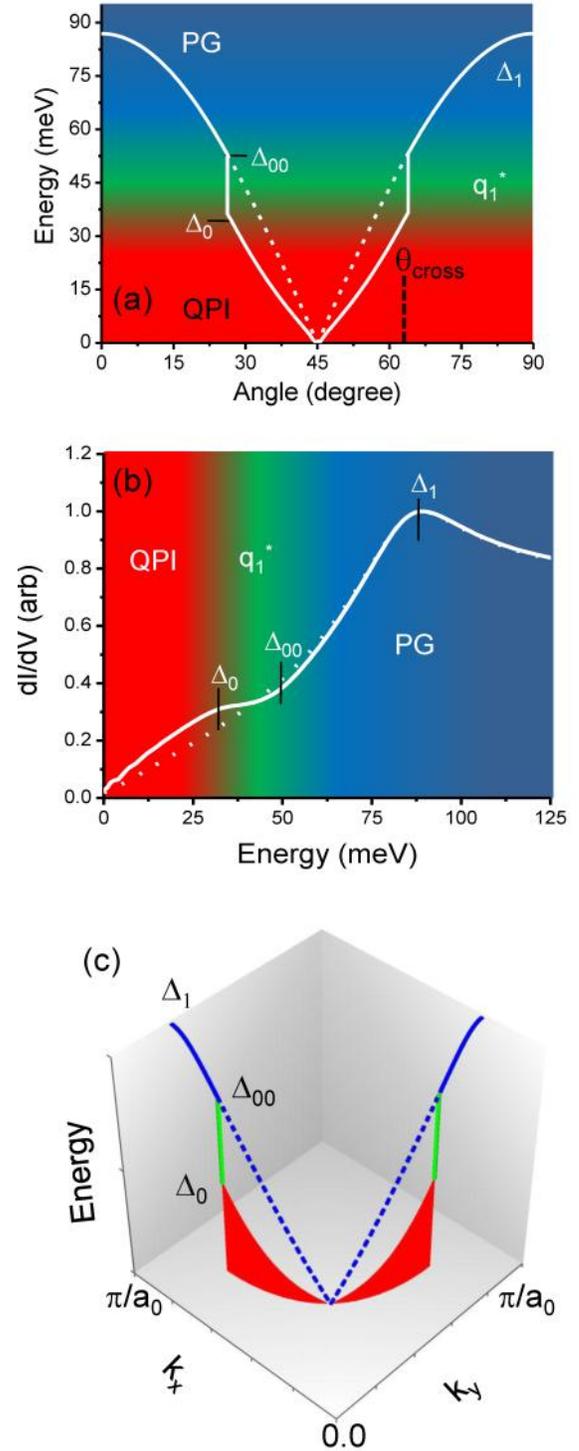

**Figure 3:** Tripartite model representations. (a) Shows the Tripartite models QPI structure. The solid line is the peak in the Tripartite $k$-space A($k$,E), while the dotted line represents the peak for an A($k$,E) with a lowest harmonic d-wave gap. At low energies, marked in red, are the coherent Bogoliubov quasiparticle excitations whose dispersion follows a d-wave gap + a higher harmonic d-wave term. These states end at $\Delta_0$ in energy or at $\theta_{\mathrm{cross}}$ in angle measured from ($\pi/a_0,\pi/a_0$) where the $q_1^*$ modulation begins (marked in green). In the model, the $q_1^*$ modulation is caused by a non-dispersive transition between the end of the dispersive QPI and the higher energy, lowest order d-wave gap of the pseudogap (marked in blue). (b) Shows the LDOS(E) for the same A($k$,E) as in (a). This LDOS(E) is calculated by integrating the A($k$,E) over the Brillion zone. The dispersive low energy states become the low energy 'V' in the LDOS, while the non-dispersive $q_1^*$ modulation region becomes the kink. The 'gap map' peak is created by the lowest order d-wave that exists in high energy, marked in blue (c) 3D $k$-space representation of the Tripartite model that is used to fit the positive energies data. This plot represents one quarter of the Brillion zone, showing the d-wave + higher harmonic in red, the non-dispersive $q_1^*$ modulation in green, and the pseudogap d-wave in blue.

The gap structure used is not the only model that can produce a spectral density of states that matches the QPI and LDOS(E) data. It is simply a model based off historical analyses of the LDOS and QPI that allows both data to be fit with the same parameters. The Tripartite gap structure also has the advantage of defining two additional energies, which have been matched to other observables[11]. These two energies are the last energy on the higher harmonic arc and the first energy on the lowest order d-wave. These are defined by:

$$\Delta_0 = \Delta_1 \cdot (B \cdot Cos[2\theta_{cross}] + (1-B) \cdot Cos[6\theta_{cross}]) \qquad 6$$

$$\Delta_{00} = \Delta_1 \cdot Cos[2\theta_{cross}] \qquad 7$$

The two energies bracket a non-dispersive region in momentum that when the A(**k**,E) is integrated, become a flat region over the energy span of $\Delta_0$ - $\Delta_{00}$ in the LDOS(E). In figure 3 this non-dispersive region is marked in green. Above this non-dispersive region in energy resides the lowest harmonic d-wave dispersion. This lowest order high energy d-wave gap is necessary in order to fit the LDOS(E) peak at $\Delta_1$ and its inclusion in the model may only be a way of representing the LDOS(E) $\Delta_1$ peak in a continuous manner. That is, it does not necessarily come from the region in momentum space outside the PAF-zone boundary. SI-STM has no **k**-space signal in that region and the location of the lowest harmonic gap structure in **k**-space makes no difference to any of the conclusions we draw in this paper.

Figure 3c presents an overview of the three regions of the phenomenological gap structure as defined by the Tripartite model in **k**-space. In red is the low energy higher harmonic dispersion of Bogoliubov quasiparticles that ends at $\theta_{cross}$. In green, a non-dispersive region that connects the low energy higher harmonic gap to a lowest order d-wave gap. In blue the lowest harmonic gap is highlighted. This Tripartite structure defines three different regions that are each associated with three different **q**-space excitations, examples of which are shown in figure 2. In red at low energies is the dispersive QPI, which are excitations from the superconducting ground state likely caused by weak scattering[5,10,32]. Above these states in energy is the non-dispersive $q_1^*$ modulation whose -3 dB intensity corresponds to the end of the kink in the mean LDOS(E)[11]. The pseudogap state, marked in blue, is the peak in the positive LDOS(E) and the corresponding peak in the pseudogap $S_{x,y}$ **q**-vectors and Bragg peaks, $Q_{x,y}$, intensity[8,9,11].

Seven parameters are used to generate the model LDOS(E) that is used to match the data. The first parameter is a factor that scales the total number of the states. This factor is set by the setup condition and is not truly a free parameter; however for ease of the fitting it is allowed to vary. Next is a linear background slope that corrects for 1$^{st}$ order background effects. The background slope is needed even though only the positive side (empty states) of the spectrum is fitted and may be a result of varying band structure or other effects[33]. There are then the three energy scales, $\Delta_0$, $\Delta_{00}$ and $\Delta_1$, along with the two lifetime terms. The $\Gamma_1$ energy constant lifetime term and the $\alpha$ term, which is used to define the liner in energy lifetime term, $\Gamma_2=\alpha$*E. There is finally, a B term that sets the amount of higher harmonic gap at energies below $\Delta_0$; however it is not a free parameter and is set uniquely by $\Delta_0$, $\Delta_{00}$ and $\Delta_1$. It is important to note that in our previous paper we have shown that each of the three energy scale can be measured independently of the Tripartite model in the mean spectra[11].

### 4. Fitting Procedure

In order to calculate the A(**k**,E) a band structure for each sample is needed. This band structure[34] is obtained by fitting a tight binding model to the **k**-space points extracted from the QPI using the octet model[15]. The A(**k**,E) is then calculated over a 128 - 512 pixel squared distorted grid with a **k** = 0 to $\pi/a_0$, for each energy in the data set. Since the A(**k**,E) is sharply peaked at the Fermi momentum, the calculation grid is distorted to decrease the spacing between points near the Fermi surface and to increase the grid spacing far away from it. This distortion gives a higher resolution where there is weight in the A(**k**,E) and low resolution where there is little or no A(**k**,E) weight. A higher resolution undistorted grid has been compared to the calculation using the distorted grid to ensure that this variable resolution spacing does not cause distortions in the LDOS(E).

To speed up the A(**k**,E) calculation and integration for LDOS(E) fitting, the calculations are carried out on a GPU running at ~ 0.5 Teraflops (actual GPU usage) and the fit is carried out using a Levenberg-Marquardt nonlinear least squares algorithum[35]. In order to ensure global minimization each LDOS(E) curve is fit 20-30 times using different starting parameters. The LDOS(E) are then refit using a higher resolution grid in **k**-space for the A(**k**,E) generation. After the initial fitting pass is complete, the results are cleaned up by refitting points that have high $\chi^2$, or that have a high departure in their parameterization from that of their surrounding points. This constraint assumes that none of the fitting parameters vary greatly (30-100% depending on the parameter) over the length scale of ~ 0.2 - 0.4 nm. This removes bad pixels in the fitting parameters, but does not change any of the observables (all features here can be seen without this process). The cleanup process allows the successfully fitting of locations where the LDOS(**r**,E) is extremely noisy or where the features are close together allowing for two radically different parameterizations to be valid that has little variation in $\chi^2$.

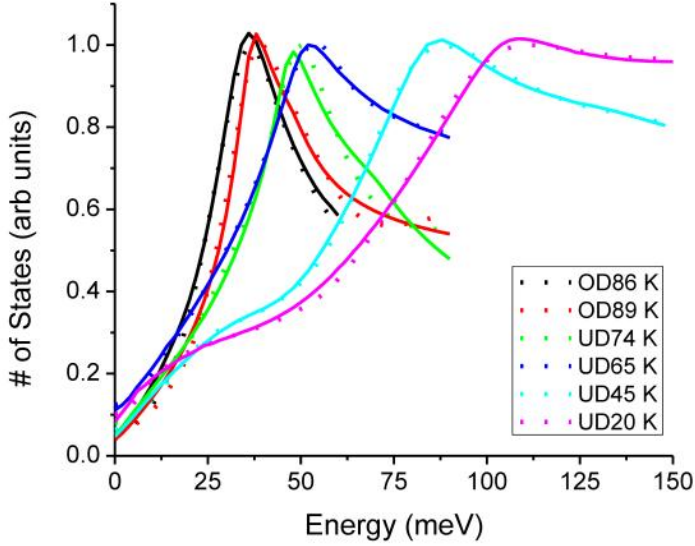

**Figure 4:** Examples of the fits for the mean $\Delta_1$ averaged LDOS(E) for each doping. The individual spectra in each dopings data set were fitted and then the data and the fits were averaged over a bin size of $\Delta_1$ +/- 2 meV for this figure. This shows the Tripartite models ability to, on average, reproduce the LDOS(E) from p ~ 0.19 to p ~ 0.06.

## 5. Fitting Results

In figure 4 the individual averaged spectra and their fits for the doping dependent samples are shown. This shows that the Tripartite model captures the majority of the LDOS(E)'s features. The model has the ability to fit the smooth evolution in the kink and peak structure, which is necessary in order to quantitatively describe the low energy positive LDOS(E)'s.

Individual doping maps revel a wide range of $\Delta_0$, $\Delta_{00}$, and $\Delta_1$'s that vary spatially. Plotted in figure 5a are examples of the sorted LDOS(E)'s of a UD45 K ($p = 0.08$) sample. In figure 5a a specific $\Delta_1$ is selected and the data is sorted to present a series of different LDOS(E)s that have a range of $\Delta_0$ for that $\Delta_1$. This sorting procedure can be applied to higher dopings where the kink begins to disappear into the $\Delta_1$ peak. In figure 5b, UD74 K ($p = 0.14$), the change in $\Delta_0$ is similar and illustrates the model and the fitting procedures ability to measure the merging of the kink and the $\Delta_1$ peak. As a further general example in figure 5c-d examples of the types of curves that exist in a 40.4 nm$^2$ field of view of the UD45 K data are plotted. This plot shows the wide range of possible LDOS(E)'s that must be and are able to be fit by the Tripartite model.

### 5a. Doping dependence of Tripartite Observables

Each energy scale shows significant spatial variation for a given doping. The histograms of their values can be compared across dopings by normalizing all three of the different energy scale maps to their mean value. Figure 6 presents these normalized histograms of the energy scales for the six dopings. Complete data sets for all dopings along with normalized $\chi^2$ values can be found in supplementary figures 1 - 10. The lower two energy scales, $\Delta_0$ and $\Delta_{00}$ define the end of the V-shaped LDOS and the end of the kink and both have similar disorder as the $\Delta_1$ energy scale for dopings near optimal ($p > 0.14$). Some of this similarity is due to the kink merging with $\Delta_1$, which sets all three energy scales to be equivalent in the model.

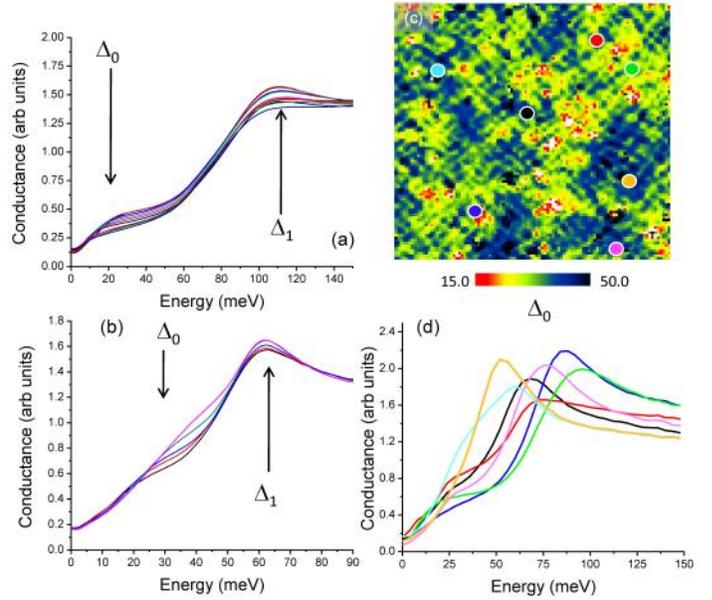

**Figure 5:** Two sets of sorted spectra from the UD45 K (a) and UD74 K (b) samples. In (a) all the data with a $\Delta_1 = 112$ meV +/- 4 meV is selected. This selected data is sorted for a range of $\Delta_0$ values from 36 meV to 12 meV and plotted. This demonstrates a range of $\Delta_0$ values for a single $\Delta_1$ value of a single data set and the Tripartite models ability to accurately measure this variation. Later in the paper this distribution is demonstrated to be due to the $q_1^*$ modulation (see section on the $q_1^*$ modulation). In (b) the same process is carried out for UD74 K at a $\Delta_1 = 62$ meV +/- 2 meV and for a range of $\Delta_0$ from 23 meV to 53 meV. (c) Is a map of $\Delta_0$ energy scales for UD45 K data, for each colored dot the extracted the LDOS(E) curve is shown in (d). There is a wide range of different positive LDOS(E) that any model must be able to fit.

The points where the kink is not observed are locations where either the phenomena of the pseudogap $\Delta_1$ peak merges with phenomena of the kink, or where the phenomena represented by the kink and the pseudogap peak overlap when measured with SI-STM and are not distinguishable in the LDOS(E)'s. The fourth column of figure 6 has $\Delta_{00} - \Delta_0$ maps for all six dopings, which separates the field of view into kinked regions and regions with no kinks. This kink map shows the kink spanning a larger energy range as well as increasing in percentage of the field of view as the doping is diminished.

The doping dependence of the three energy scale maps shows that their spatial patterns of the three scales are linked. All four columns have the same boundary structure/length scale and have high normalized cross correlation coefficients between the maps (table 1). The full width half maximum of the autocorrelation of the energy scale maps is constant across all dopings (table 2) within the limited spatial resolution of this study. This demonstrates that the length scale and disorder is independent of doping.

In the far underdoped regime ($p < 0.10$), $\Delta_0$ and $\Delta_{00}$ are modulated by the $q_1^*$ modulation pattern; however we will show that this is a likely artifact of the $q_1^*$ modulations signature, rather than an actual modulation of energy scales (see section: $q_1^*$ modulation of LDOS($r$,E) ). The correlation between the lowest energy scale, $\Delta_0$, and the pseudogap energy scale, $\Delta_1$, changes from a positive correlation to a negative one when $\Delta_1$ passes ~ 60 meV or a corresponding doping of approximately $p = 0.10$. If $\Delta_0$ is the energy of the superconducting phenomena, then this is an expected change in the sign of the correlation. As the doping diminishes, the superconducting energy scale should track $T_c$ and go towards zero. However despite this change, the disorder patterns boundaries still have the same length scale and structure across all dopings (see tables 1,2).

| $p$ | 0.19 | 0.17 | 0.14 | 0.10 | 0.08 | 0.06 |
|---|---|---|---|---|---|---|
| $\Delta_1 * \Delta_{00}$ | 0.93 | 0.94 | 0.87 | 0.66 | 0.55 | 0.60 |
| $\Delta_1 * \Delta_0$ | 0.87 | 0.91 | 0.45 | 0.36 | -0.43 | -0.40 |
| $\Delta_1 * (\Delta_{00} - \Delta_0)$ | 0.19 | 0.40 | 0.75 | 0.47 | 0.81 | 0.84 |

**Table 1:** The normalized cross correlations for the data shown in figure 6. As the kink goes to zero in a field of view, the normalized cross correlation of its span ($\Delta_1 * (\Delta_{00} - \Delta_0)$) goes to zero ($p = 0.19, 0.17$). While the spatial patterns in figure 6 are all similar, the change in relationship between $\Delta_0$ and $\Delta_1$ causes a change in the normalized cross correlation at low dopings as $\Delta_0$ goes towards zero.

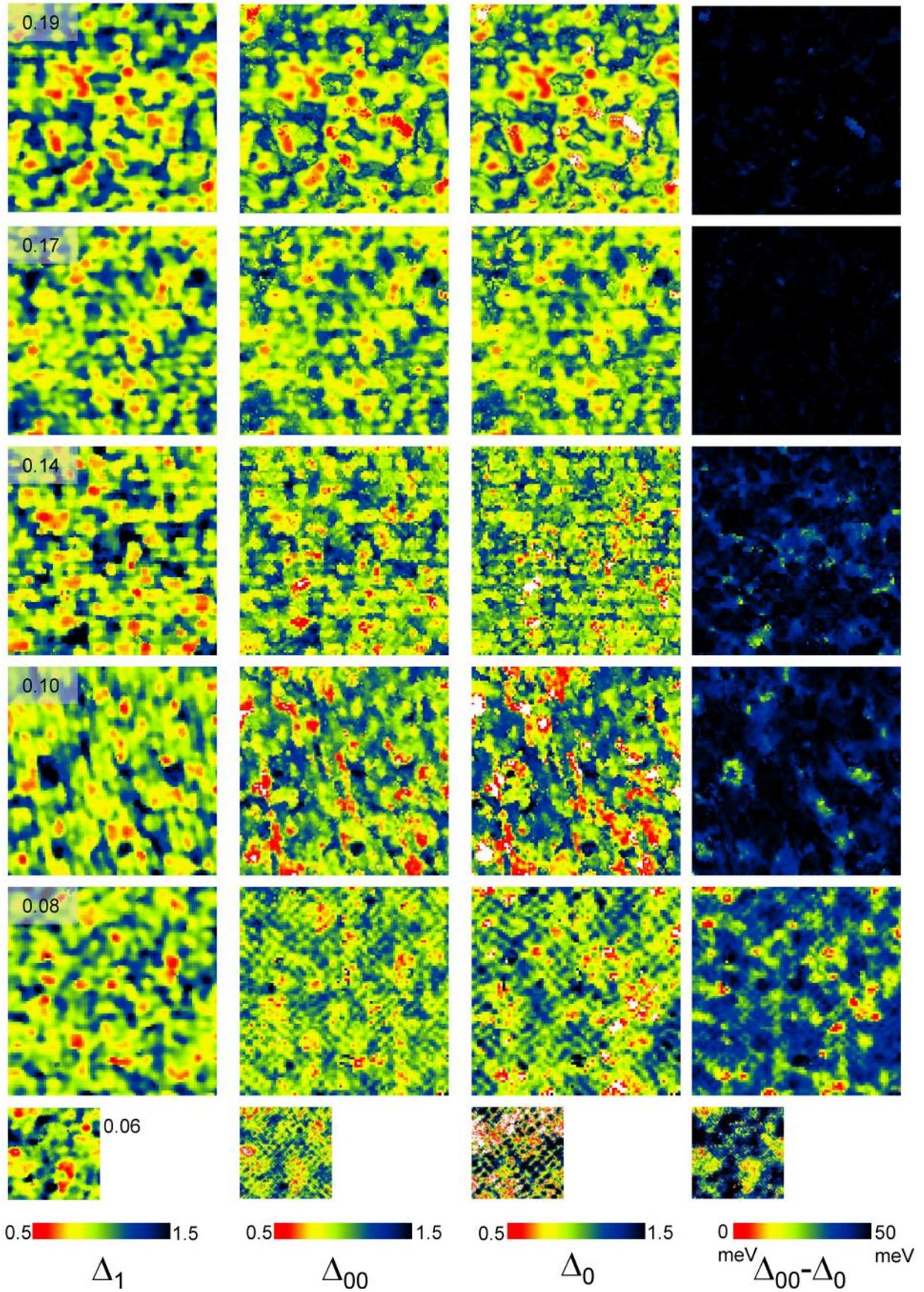

**Figure 6:** The three energy scales $\Delta_1$, $\Delta_{00}$, $\Delta_0$ for dopings 0.19 - 0.06 normalized to the average value of each at that doping. This normalization highlights the similarities that exist in the spatial patterns between dopings and energy scales. The fourth column, $\Delta_{00} - \Delta_0$, represents the kink width as a function of doping. All the data sets have an identical field of view of 40.4 nm$^2$ except for the $p$ = 0.06 data set which has a 20.2 nm$^2$ field of view. This figure shows that the energy scales have the same spatial pattern for all dopings. At lower dopings the $q_1^*$ modulation can be seen in the fitted $\Delta_0$ and $\Delta_{00}$ values. This is most likely an artifact due to the $q_1^*$ modulations LDOS(E) signature (see figure 15).

| $p$ | $\Delta_1$ | $\Delta_{00}$ | $\Delta_0$ |
|---|---|---|---|
| 0.19 | 2.2 nm | 2.2 nm | 2.3 nm |
| 0.17 | 2.4 nm | 2.3 nm | 2.3 nm |
| 0.14 | 2.3 nm | 2.0 nm | 1.7 nm |
| 0.10 | 2.9 nm | 2.3 nm | 2.4 nm |
| 0.08 | 2.8 nm | 2.5 nm | 3.0 nm |
| 0.06 | 2.4 nm | 2.4 nm | 2.0 nm |

**Table 2:** Full width half maximum (FWHM) for the autocorrelations (0,0) peaks for all the energy scales. The autocorrelations (0,0) FWHM gives the effective length scale of an image. The line cuts through the autocorrelation peak are shown in supplementary figure 11. The error in the FWHM is set by the resolution of the data, which for this study ranges from 0.5 nm to 0.2 nm over the data sets. This shows there is no variation of the length scale for any of the three energy scales.

### 5b. Small Field of View and $\Delta_1$ Gradient

In figure 7 a 15.4 nm$^2$ field of view of a UD74 K sample has been fitted. The resulting parameterization shows that all three energy scales are disordered even on small length scales[14]. This small field of view data set has a higher overall background noise due to the reduced time over which the data was taken and the increased data points per curve. Despite this noise the data provides a clear picture of the disorder on the lattice length scale. In figure 7 the $\Delta_1$ map shows the familiar gap map pattern[14] as well as atomic modulations that are present in all high resolution data sets. The $\Delta_1$ map has the standard length scale of 2 - 4 nm patch work that has been seen universally for all dopings of $Bi_2Sr_2CaCu_2O_{8+x}$ in the past[13,14]. The energy where the kink begins, $\Delta_0$, has a low normalized cross correlation with $\Delta_1$ of ~ 0.3; however, the $\Delta_0$ map does have an underlying structure that matches the $\Delta_1$ disorder. The ending of the kink, $\Delta_{00}$, has more in common with $\Delta_1$ and has a cross correlation coefficient of 0.65. This is in part due to $\Delta_{00}$ being set equal to $\Delta_1$ where there is no kink in the field of view.

The energy dependent lifetime term, $\alpha$, has a positive correlation with $\Delta_1$ and it is higher in value at the edges of the $\Delta_1$ patches. This edge effect is highlighted by comparing the $\alpha$ map to the gradient of $\Delta_1$ which outlines the individual patches of the $\Delta_1$ disorder. In figure 7f a clear correlation between $\alpha$ and Grad[$\Delta_1$] is displayed. $\alpha$ resembles $\Delta_1$'s gradient more than it does $\Delta_1$ itself. At this doping and there is a normalized cross correlation coefficient of 0.5 between $\alpha$ and Grad[$\Delta_1$], which is bigger than the normalized cross correlation coefficient between $\alpha$ and $\Delta_1$ of 0.4. While this effect could be an artifact caused by a blunt SI-STM tip, which would have the effect of blurring the distinctions between the patches causing an effective increase in $\alpha$, we know it is not. This is because as figure 7e demonstrates that there is a single sharp tip for this data set showing that either the $\Delta_1$ patch is the source of scattering that generates the $\alpha$ disorder or that the $\Delta_1$ patches are confined by doping independent, weak scatterers that result in decreased lifetimes for the pseudogap/superconducting gap. These scatterers would be inherent to the crystal structure and would act to define the patchwork.

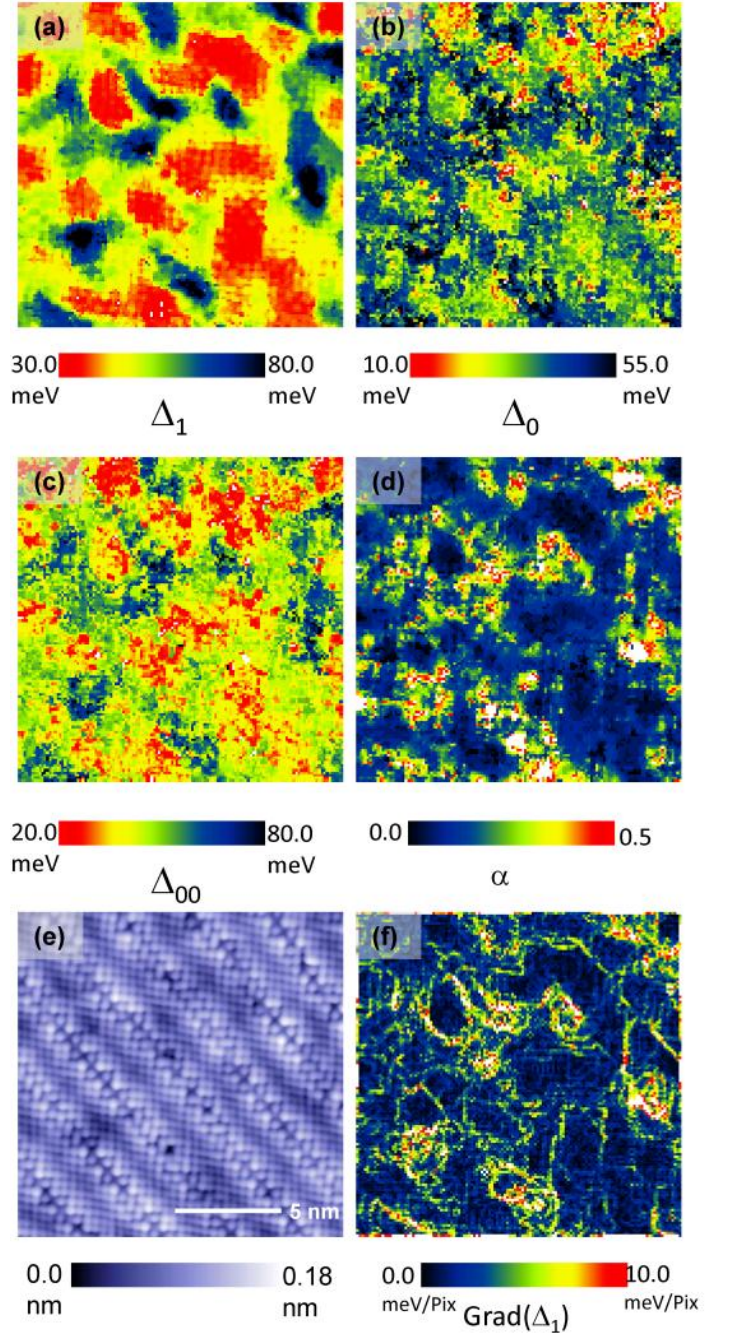

**Figure 7:** The fitting parameters for a 15.4 nm$^2$ field of view, with a doping of $p \sim 0.14$, and a $T_c$ of UD74 K. (a) shows the high energy LDOS(E) peak map of the sample ($\Delta_1$), while (b) is the lower energy scale, $\Delta_0$ which is has a normalized cross correlation coefficient with $\Delta_1$ of 0.35. $\Delta_0$ represents the beginning of the kink energy and the end of the dispersive QPI. (c) Shows the end of the kink energy scale, $\Delta_{00}$, which has a stronger correlation with $\Delta_1$ of 0.66. The high correlation is partially the result of $\Delta_{00}$ not being fully differentiated from $\Delta_1$ at this high doping. (d) Shows the linear in energy scattering term, $\alpha$, that has a normalized cross correlation with $\Delta_1$ of 0.41. In (e) the simultaneous topograph demonstrates a clean single tip with atomic resolution. (f) Shows the gradient of the $\Delta_1$ map, (a), revealing that $\alpha$ has a strong normalized cross correlation with the Grad[$\Delta_1$] of 0.46.

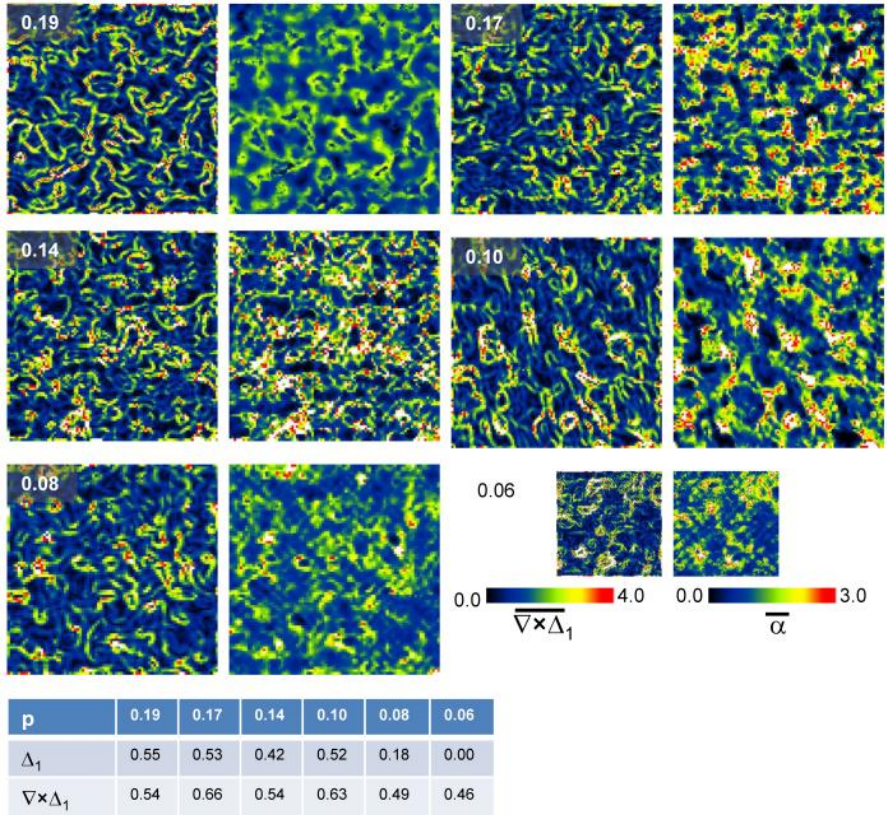

| p | 0.19 | 0.17 | 0.14 | 0.10 | 0.08 | 0.06 |
|---|------|------|------|------|------|------|
| $\Delta_1$ | 0.55 | 0.53 | 0.42 | 0.52 | 0.18 | 0.00 |
| $\nabla \times \Delta_1$ | 0.54 | 0.66 | 0.54 | 0.63 | 0.49 | 0.46 |

**Figure 8:** Grad[$\Delta_1$] compared to the local lifetime variations over a doping range of 0.19 - 0.06. The two columns show that the $\alpha$ variation follows the Grad[$\Delta_1$] for all dopings. The table in the figure lists the normalized cross correlation coefficients for $\alpha$ with $\Delta_1$ and $\alpha$ cross correlated with Grad[$\Delta_1$]. This figure demonstrates that every doping has a low lifetime at the boundaries between the patches.

## 5c. Short Quasiparticle Lifetimes at Patch Boundaries

Figure 8 shows the relationship between $\alpha$, $\Delta_1$, and Grad[$\Delta_1$] for all the dopings. In the accompanying table, the normalized cross correlations between $\alpha$, $\Delta_1$ and Grad[$\Delta_1$] are shown. Previously[13] it has been reported that $\alpha$ has a linear dependence on $\Delta_1$ value for dopings above 0.10 and in the table in figure 8, the positive normalized cross correlations between $\Delta_1$ and $\alpha$ at dopings above 0.10 supports this. At dopings near optimal ($p$ = 0.19, 0.17) the $\alpha$ maps have a high normalized cross correlation coefficient with the $\Delta_1$ maps and the Grad[$\Delta_1$] maps. As the doping levels are decreased, the correlation between $\alpha$ and $\Delta_1$ disappears. This can be seen both visually in figure 8 and in the values for the normalized cross correlation coefficients present in the table. However, the normalized cross correlation coefficients remain high between Grad[$\Delta_1$] and $\alpha$, even for dopings below 0.10 where the normalized cross correlation between $\Delta_1$ and $\alpha$ reaches zero. The drop in the correlation between $\alpha$ and $\Delta_1$ marks a change in the low energy electronic structure of the empty states LDOS(E).

In figure 9 the average value of $\alpha$ as a function of $\Delta_1$ is plotted. For optimal to moderately underdoped samples ($\Delta_1$ = 30 - 60 meV) there is a linear scaling of $\alpha$ with $\Delta_1$. However, above $\Delta_1 \sim$ 70 meV there is a decrease in the value of $\alpha$ and above $\Delta_1 \sim$ 80 meV, $\alpha$ has no or a negative dependence on $\Delta_1$. These high $\Delta_1$ regions correspond to dopings where there is a low (< 0.20) normalized cross correlation coefficient between $\alpha$ and $\Delta_1$. This represents a change in the scattering dependence for the underdoped samples. The reason behind this shift is not clear since there are several possible origins of the linear in lifetime scattering rate[21], including elastic weak scattering by out of plane dopants[20], elastic pairing disorder scattering by the dopants[36], or inelastic scattering by spin fluctuations[21].

A recent neutron scattering study[37] has detected a $q$ = 0 (or $Q_{x,y}$) spin structure in the pseudogap state that could correspond to the excitations observed at E = $\Delta_1$ by SI-STM[4,8,9]. There are also other proposed smectic spin stripes that could correspond to the pseudogaps $S_{x,y}$[38]. If the pseudogap nematic states do represent a spin ordered state, a possibility considering neutron scattering[39,40] and SI-STM results[8,9,12], then the onset of this state could dramatically change the scattering rate.

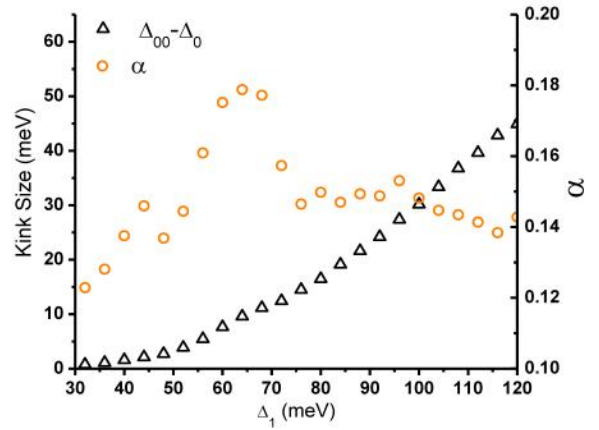

**Figure 9:** Plots the kink size ($\Delta_{00} - \Delta_0$) on the left axis as open black triangles and the energy dependent scattering rate, $\alpha$ on the right axis as open red circles. $\alpha$ increases linearly until $\Delta_1 \sim$ 70 meV. It then decreases in value and becomes independent of $\Delta_1$. The kink increases in spain constantly as $\Delta_1$ increases. Since the transition from a linear increasing $\alpha$ to an $\alpha$ that is independent of $\Delta_1$ occurs when the kink is large, it is unlikely that this change in $\alpha$s dependence is due to an error in the fitting.

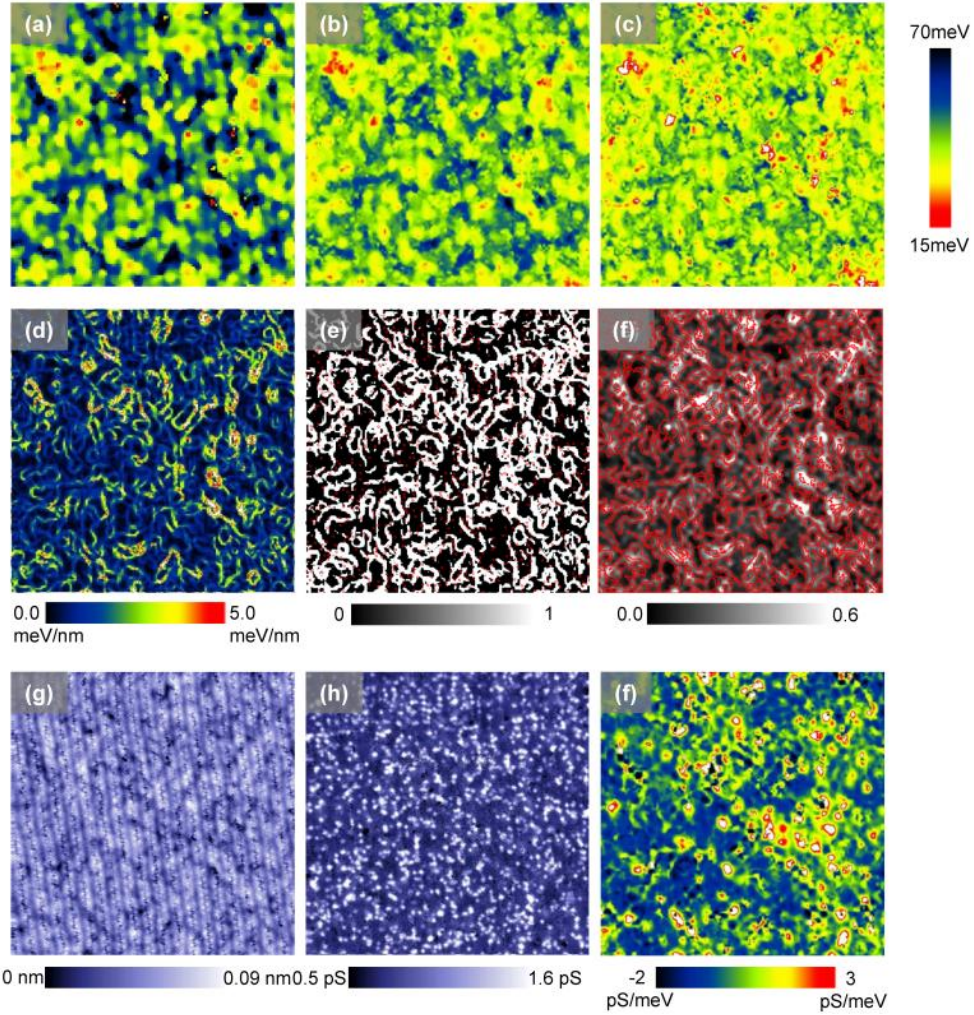

**Figure 10:** Oxygen doping atoms relationship to the Tripartite model parameters for a 48 nm$^2$ field of view. (a) $\Delta_1$, (b) $\Delta_{00}$, (c) $\Delta_0$, (d) Grad[$\Delta_1$], (e) A binary mask of Grad[$\Delta_1$]/$\Delta_1$ with oxygen positions overlaid as red circles, this demonstrates that the oxygen atoms do not sit preferentially on the boundaries between the regions, and therefore are not the cause of the regions or of the low lifetime at their intersection. (f) is the corresponding $\alpha$ map with contours of (e) overlaid in red. (g) is the simultaneous topograph and (h) the oxygen map dI/dV at E = - 960 meV. (g) is the empty states background slope. The normalized cross correlation coefficient between (d) the Grad[$\Delta_1$] and (f) $\alpha$ is 0.56. The normalized cross correlation coefficient between $\Delta_1$ and oxygen map is 0.29, between Grad[$\Delta_1$] and oxygen map, 0.122, between $\Delta_{00}$ and oxygen map, 0.24, and for $\Delta_0$ and oxygen map, 0.131.

### 5d. Oxygen Dopant Atoms and Tripartite Observables

One of the mechanisms that has been reported[22,24] to drive the disorder in the electronic structure is the inhomogeneous spatial distribution of the dopant oxygen atoms. These oxygen atoms are spatially mapped by measuring a broad peak in the differential tunneling current that occurs below -700 meV in energy and has a maximum at -960 meV[22] or by measuring a peak at high positive energies[24]. In this study we only have access to the negative energy oxygen data so we must limit our analysis to that. For both positive and negative energy peaks, the dopant oxygen atoms tend to be found in the regions where $\Delta_1$ is large compared to the <$\Delta_1$> for that doping. The dopant oxygen's peak has a normalized cross correlation with the $\Delta_1$ map of ~ 0.3 for the negative energy peaks and ~ 0.6 for the positive energy peaks. Comparing the placement of the negative bias oxygen atoms to the Tripartite spatially resolved parameterization of the LDOS(E) shows the effect that dopant atoms have on the energy scales and the lifetimes. Figure 10 shows the results of this analysis for a moderately underdoped sample with nickel impurities ($\Delta_1$ average of 55 meV corresponding to a *p* of 0.14). The nickel impurities[41] have LDOS(E) peaks at ~10-20 meV and tend to suppress the $\Delta_1$ peak locally, this drives up the error in our fits (see supplementary figure 10). However we are still able to successfully fit the majority of the sample. In this sample there exists the same $\Delta_1$, $\Delta_{00}$, $\Delta_0$ universal patterns and all three energy scales are positively correlated with the oxygen map. Each energy scale has similar normalized cross correlation coefficients as reported previously for $\Delta_1$ (0.29 for $\Delta_1$, 0.24 for $\Delta_{00}$, and 0.13 for $\Delta_0$).

The $\alpha$ map is correlated with the Grad[$\Delta_1$] map and has a normalized cross correlation coefficient of 0.56. By calculating the Grad[$\Delta_1$]/$\Delta_1$ and masking it into a binary valued boundary map, the number of oxygen atoms that reside along the boundaries between patches can be counted. This analysis, shown in figure 10e, demonstrates that 43% of the oxygen dopant atoms lie along the patch boundaries. However, the patch boundary mask created has 40% of the field of view with a value of 1. Randomly rotating or shifting the patch boundary map results in 40% of the oxygen atoms lying along the patch boundaries. Therefore, the oxygen atoms do not define the patch boundaries. For the same patch boundary mask the average $\alpha$ value of the pixels on the boundary is 0.26, which is larger than the average α for the unmasked field of view, 0.19, showing that the scattering does increases at the boundaries. In figure 10f this relationship is highlighted by overlaying the outline of the patch boundary mask from figure 10e, with the α map. The outline follows the regions of high α almost exactly. The high α at the boundaries and the lack of an oxygen correlation shows that there is some other mechanism creating the patchwork structure. The positive normalized cross correlation between the oxygen dopant atoms and both the $\Delta_1$ map and the α map shows that the oxygen atoms do have an effect on the value of both the scattering and the $\Delta_1$ gap locally. We have seen that the normalized cross correlation between $\Delta_1$ and α disappears at low dopings. This could be due to the decrease in the number of dopant atoms causing a decrease in the scattering that is correlated with $\Delta_1$. This would leave only the scattering that is correlated with the patchwork boundaries behind and is consistent with the theory that the local $\Delta_1$ value is driven by local exchange or superexchange enhancement[23,42,43].

The background linear slope term has a normalized cross correlation with the oxygen map of 0.25, which is similar in value to that of the normalized cross correlation coefficient between the oxygen map and $\Delta_1$. The background slope has no other normalized cross correlation coefficients of significance. While the spatial structure of the empty states slope map remains consistent across dopings and samples, the maximum slope and the mean slope value vary randomly from sample to sample. This suggests that there is a dependence on the setup condition for the scale and average of this slope. This empty states slope either measures some local variation in that setup condition that is strongly influenced by the oxygen atoms, or the slope is a remnant of the higher energy positive states of the some of the oxygen atoms[24]. Since the positive oxygen feature has some spatial overlap with the negative oxygen peak positions[24], this could explain the correlation. The slopes for all the data sets can be found in supplementary figures 1-9. As a practical matter, this correlation could represent a way to image the oxygen atoms without having to resort to high energy scans which are often destructive to the tip and the sample.

In figure 11 the fitting parameter values are plotted as a function of distance away from the oxygen atoms. Since the oxygen atoms map consists of small area peaks and all the other observable maps are composed of larger, extended regions, this analysis allows a clearer determination of the relationships between the oxygen atoms position and the Tripartite fit parameters. All three energy scales ($\Delta_1$, $\Delta_{00}$, $\Delta_0$) decrease with increasing distance from the oxygen atoms. The positive energy background slope is highly dependent on the oxygen atoms, falling off to the background level within 1 nm, confirming the influence of the oxygen atoms on the slope. α shows a decrease in its value with increasing distance from the oxygen atoms over a distance comparable to the patch sizes. This is consistent with previous observations that the oxygen tend to be found in the higher $\Delta_1$, suppressed peak regions of the field of view[22]. However the change in the number of the oxygen dopant atoms with doping compared to the universal nature of the patchwork disorder (figure 6) along with the fact that the oxygen dopant atoms do not lie on the patch boundaries, proves that the oxygen atoms do not define the structure of the local disorder. Instead they influence the $\Delta_1$ value and α value inside a separately defined patchwork structure.

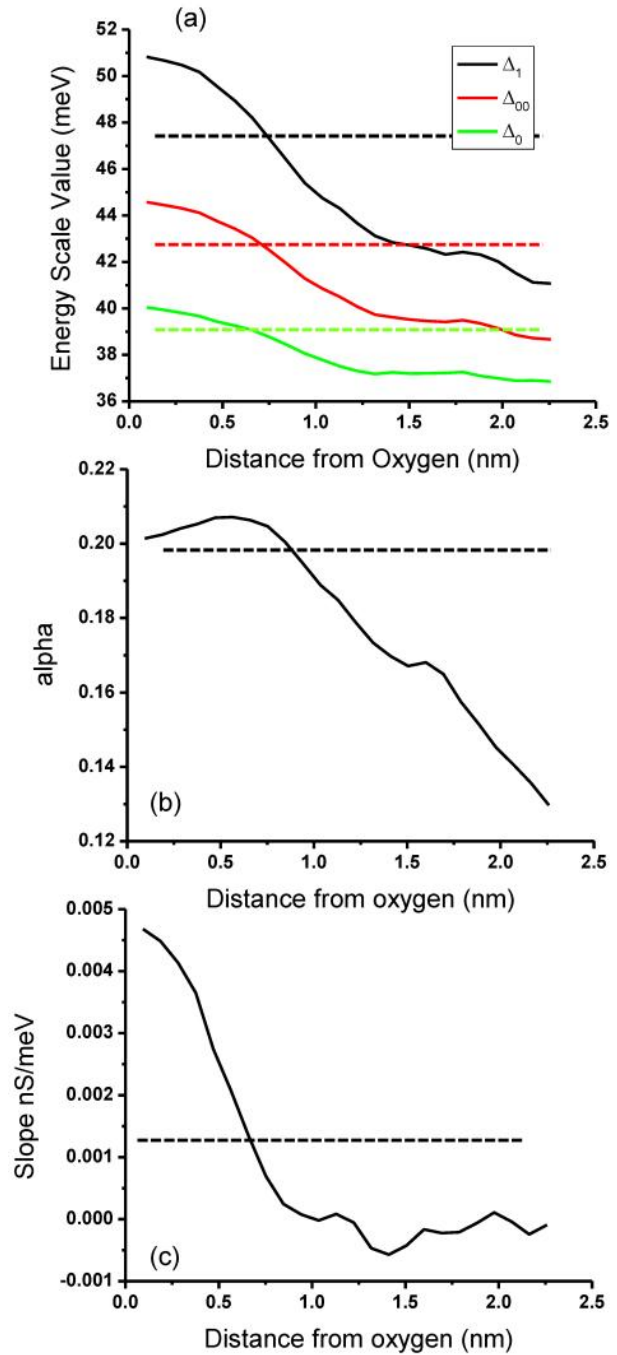

**Figure 11:** The Tripartite model parameters as a function of distance from the oxygen atoms using the data in figure 15. The dotted lines are the average value of that parameter for the field of view. In (a), the three energy scales are plotted as a function of distance from the oxygen atoms. (b) α as a function of distance from the oxygen atoms. α decreases in value over a length scale consistent with the disorder patch size. (c) The positive energy background slope, which falls to zero within 1 nm from the oxygen atoms.

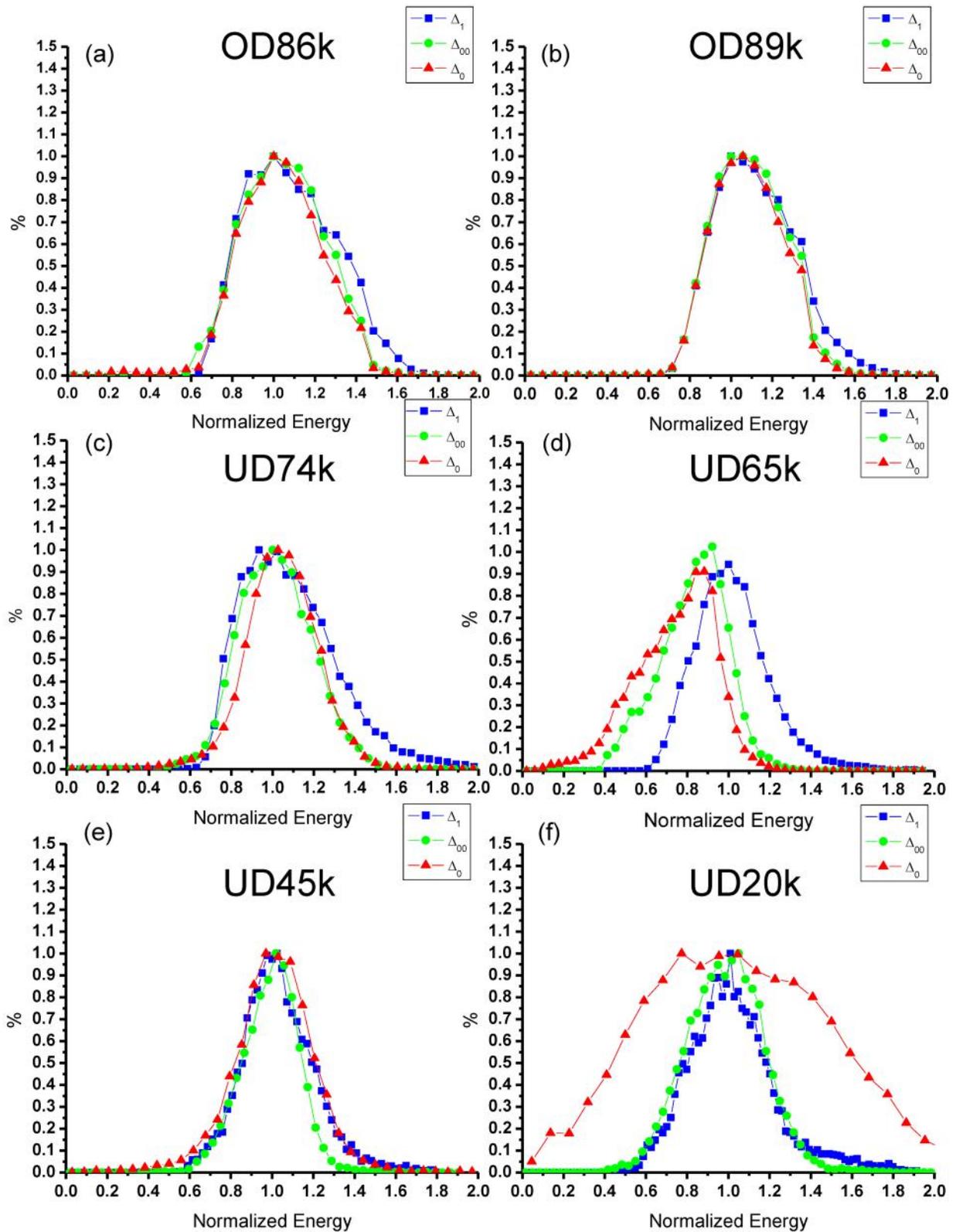

**Figure 12:** The normalized distributions of the three energy scales for all six samples. Since all the parameters have the same normalized distributions and patch patterns/length scales, they are most likely all related.

## 5e. Single Parameter Driven Doping Dependence of Length and Energy Scales

As the doping changes, the length scale of the disorder remains the same (see table 2 and figure 6). In samples with dopings at and underneath $p = 0.08$, dysprosium dopants are used to achieve a lower doping level and there addition also does not affect these length scales. There is a positive correlation between the $\Delta_1$ disorder and the oxygen dopant atoms[22,24] and while this might pin the disorder in various configurations and set the local $\Delta_1$ values relative to the mean, the change in the number of dopant atoms by a factor of three does not affect the overall disorder length scale or pattern ($p = 0.19$ to $p = 0.06$). The fact that an identical disorder is seen for the superconducting energy ($\Delta_0$), the $q_1^*$ modulation termination ($\Delta_{00}$) and the pseudogap energy ($\Delta_1$), shows that these three phenomena are affected by the underlying disorder mechanism that is independent of the doping.

By plotting the normalized distributions of all of the energy scales, the connection between them can be demonstrated quantitatively. If there is a single valued relation between the three energy values, then their full width half maximum (FWHM) of their distributions should be identical. Figure 12 shows that the normalized distributions of the three energy scales match for almost every data set, each one having the same FWHM. The exception is the extremely underdoped UD20 K sample where the distribution of $\Delta_0$ is much larger than the distributions for $\Delta_1$ or $\Delta_{00}$. One possible explanation is that as the doping level is decreased, there must come a point where there will be regions that would not be superconducting if isolated. At these dopings we will begin to see the breakdown of the global superconductivity and the sample will need to be treated as a matrix of superconductivity patches supporting proximity superconductivity in their neighboring non-superconducting patches. The single value relationship between $\Delta_0$, $\Delta_{00}$ and $\Delta_1$ would no longer hold since proximity superconducting effects would be present. There is evidence of kink like structures in classical BCS proximity effects[44] and that could explain the different $\Delta_0$ distribution at $p = 0.06$. It is also possible that the change in FWHM is due to increased out of plane disorder caused by the low doping[45] or the decrease in screening length with low doping[46].

The Tripartite model allows the mapping of the relationship between $\Delta_0$, $\Delta_{00}$, and $\Delta_1$. It can also be used to determine that there is a single value relationship between $\Delta_1$, $\Delta_{00}$ and $\Delta_0$. In figure 13a the three energy scales with their FWHM distributions (error bars), are plotted as a function of $\Delta_1$. In figure 13, as the pseudogap energy scale, $\Delta_1$, becomes smaller, the kink and superconductive energy scale merge or overlap. As the pseudogap energy scale grows in energy, both $\Delta_{00}$ and $\Delta_0$ break away from it. At the doping where the energy scales diverge, $\Delta_{00}$ increases with a shallower slope then $\Delta_1$. $\Delta_0$'s slope, turns over and heads towards zero, bottoming out around $\Delta_1 \sim 140$ meV. Figure 13 represents the data for 6 different dopings showing clearly that there is a continuous single valued evolution in the energy scales. To infer a doping dependence, the average $\Delta_1$ value is determined from the area averaged LDOS(E) for each doping and the dispersive QPI data. These values are then fit to a polynomial and used to convert local $\Delta_1$ values into doping values. This inferred doping dependence is plotted in figure 13b and shows both a pseudogap energy scale that increases with decreasing doping and a superconducting energy scale that forms a dome. This continuous doping figure matches previous measurements of the three energy scales for discrete dopings[11] and follows a trend similar to other probes[18]. Included on the right hand axis is a temperature scale based off $8k_BT = 2\Delta$, which has been used in cuprates previously[47]. This gives the approximant temperature dependence of the energy scales.

The equal Gaussian distributions of the parameters, the continuous evolution of $\Delta_0$ and $\Delta_{00}$ with $\Delta_1$ or inferred doping and the independence of the spatial patch size from doping, all point to a single underlying parameter that varies spatially. This parameters distribution and length scale do not change with doping, but instead there is a shift in its overall value. The sample consists of patches with different values of this parameter bounded by low lifetimes where the patches intersect. In the next section we will show that the spatial excitations (QPI, $q_1^*$ modulation, pseudogap spatial pattern) are tied to these three LDOS(E) energy scales and the underlying single valued parameter and.

| $p$ | 0.19 | 0.17 | 0.14 | 0.10 | 0.08 | 0.06 |
|---|---|---|---|---|---|---|
| $T_c$ | OD 86 K | OD89 K | UD74 K | UD65 K | UD45 K | UD20 K |
| $\langle\Delta_1\rangle$ | 36 meV | 39 meV | 50 meV | 52 meV | 86 meV | 106 meV |
| $\langle\Delta_{00}\rangle$ | 35 meV | 38 meV | 44 meV | 43 meV | 52 meV | 59 meV |
| $\langle\Delta_0\rangle$ | 34 meV | 38 meV | 40 meV | 38 meV | 34 meV | 24 meV |

**Table 3:** Average energy scale values for all 6 samples.

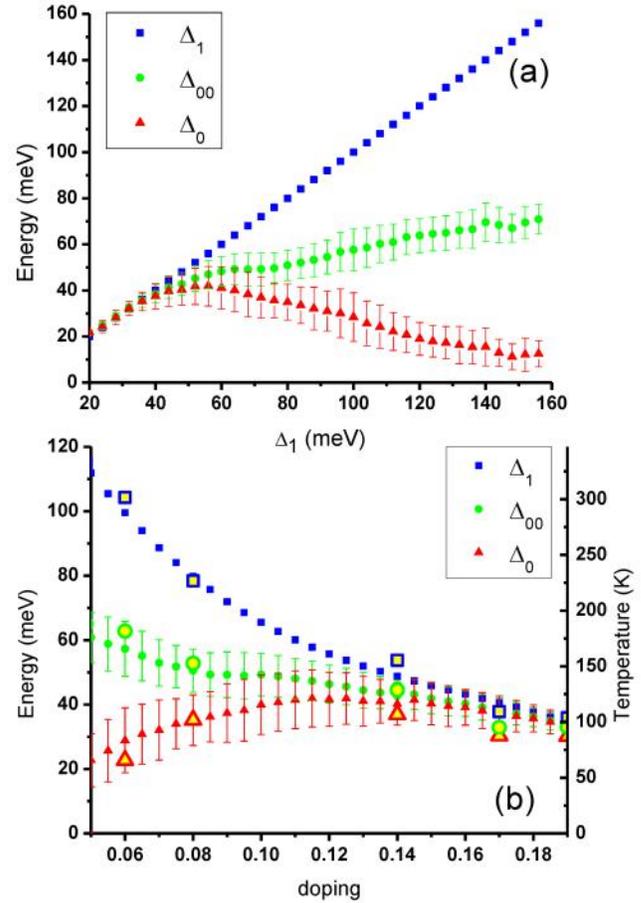

**Figure 13:** (a) The three energy scales as a function of the pseudogap energy scale, $\Delta_1$. The error bars represent the FHWM of the distribution of data point for that value of $\Delta_1$ +/- 2 meV. Using $\Delta_1$ as a measure of doping based off a polynomial fit of the $\langle\Delta_1\rangle$ for each of the six samples, (a) is mapped to doping in (b). The yellow filled larger symbols mark the mean doping values[11] in (b). The underlying parameter that defines all three energy scales is related to doping. This graph clearly shows the one parameter disorder of the three energy scales.

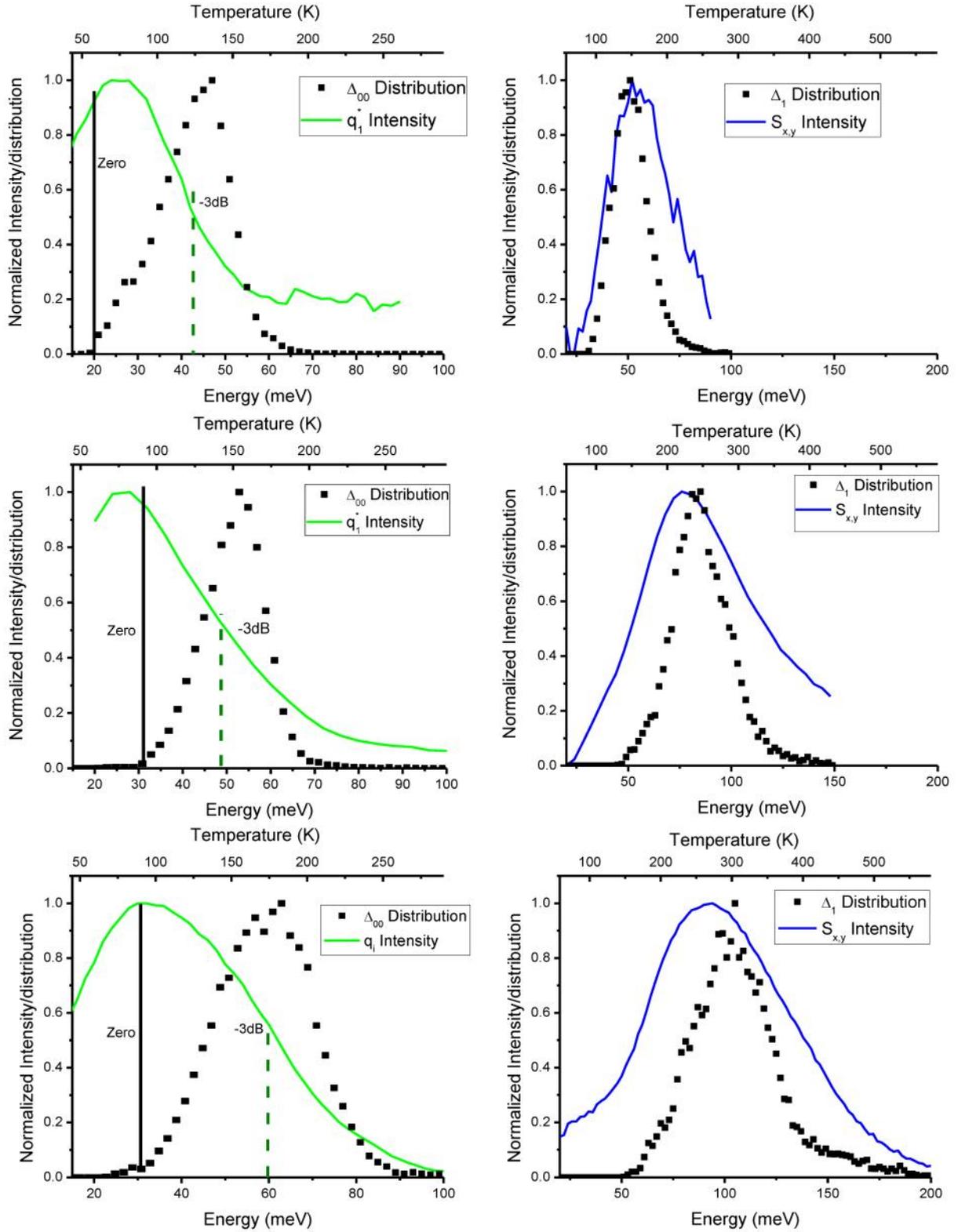

**Figure 14:** The local destruction of the $q_1^*$ modulation and the local onset of the pseudogap. $T_c$ = UD65 K, UD45 K, UD20 K data are plotted and the left hand column shows $q_1^*$ intensity overlaid with the $\Delta_{00}$ distribution. The solid black line represents the energy where none of the field of view has passed $\Delta_{00}$. The green dotted line represents the -3 dB point for the intensity of $q_1^*$. The right hand column shows the data for the $S_{x,y}$ intensity overlaid with the $\Delta_1$ distribution. We have included temperature scale here and as temperature is increased, we would expect the intensity of $q_1^*$ and $S_{x,y}$ to change accordingly. There should be gradual temperature reduction in the intensity of the ***q***-vectors due to the energy distribution of their energy ranges/peaks. For both phenomena a temperature scale follows the relation $2*E=8*k_B*T$ which has been used in previous analysis[47] of $Bi_2Sr_2CaCu_2O_{8+x}$ data.

## 6. Spatial Modulations and *k*-space Phenomena, Relations to Local Observables

There are at least three distinct spatial modulations that are present in the Fourier transform of 50 nm$^2$ field of view SI-STM data. These spatial modulations energy ranges can be determined by measurements of each of the spatial modulations different *q*-vector intensities. The spatial modulations have been matched to features in the mean LDOS(E) and their associated Tripartite energy scales[11]. The dispersive Bogoliubov QPI has been shown[4,11] to terminate at $<\Delta_0>$, the static $q_1^*$ modulation, which onsets at the QPI termination in the Z-map, declines to half its maximum amplitude at $<\Delta_{00}>$, and the pseudogap long wavelength modulation pattern[7–9,12], $S_{x,y}$, has a maximum intensity at $<\Delta_1>$. There are two options for the relation between the patches energy scales and these spatial phenomena. Either the phenomena are destroyed or created locally at the energy transitions in each patch, or their intensity is set globally by the average energy of the transitions across all the patches. In order to differentiate between the two scenarios the distribution of the local energy transitions is compared to the spatial modulations *q*-vector(s) energy dependent intensity. If the spatial modulations are locally linked to the patches energy scales, then the amount of the field of view at or past an energy scale will set the phenomena's *q*-vector intensity. If they are linked to the global average of the patches energy scale, then there will be a set percentage of the field of view that will set the termination of the phenomena depending on the strength of the phenomena.

### 6a. Spatial Modulations and $\Delta_0$, $\Delta_{00}$ and $\Delta_1$

The termination of the coherent low energy octet QPI pattern has been studied extensively[4,5,10]. It has been shown the intensities of the majority of the dispersive *q*-vectors go to zero as $<\Delta_0>$ is approched[4,5,11] and that $<\Delta_0>$ can be measured as the last energy where there is enough intrinsic signal to noise (not limited by measurement system noise, since the noise spectrum is guassian[5]) to measure the *q*-vectors and determine a *k*-space origin of the scattering[4,10,11]. This termination point occurs in momentum space at the PAF-zone boundary. The local fitting and previous *k*-space Tripartite fitting[11] results show that the energy where this intersection occurs, $<\Delta_0>$, is the average value of a distribution of the beginning of the kink (see table 3). However a detailed analysis of the QPI intensity and its relationship to the $\Delta_0$ distribution is beyond the scope of this paper, due to the complicated problem of quantifying the intensity of the dispersive peaks, taking into account setup conditions.

If the onset and termination of the $q_1^*$ modulation and the pseudogap vary in energy between the different patches, then the distribution of $\Delta_{00}$ and $\Delta_1$ will be related to $q_1^*$ and $S_{x,y}$ respectively. In the case of the $q_1^*$ modulation, if the modulation exists locally in the patch then the intensity of $q_1^*$ at a given energy will be proportional to the percentage of the field of view that has a $\Delta_{00}$ smaller than that energy. When the energy is below the lowest value in $\Delta_{00}$'s distribution, $q_1^*$ will be at its maximum, when the energy is above $<\Delta_{00}>$, $q_1^*$ will be at half its maximum and when the energy is above all the $\Delta_{00}$'s in the distribution, then $q_1^*$ intensity will be zero. The left column in figure 14 shows that this relationship is true for the three lowest dopings, which have the strongest $q_1^*$ intensities. The distribution of $\Delta_{00}$ is zero near the peak in $q_1^*$. The maximum in the Gaussian distribution of $\Delta_{00}$ sits at the -3 dB point of $q_1^*$'s intensity, and the intensity of $q_1^*$ goes to zero as the end of the $\Delta_{00}$ distribution is reached. This is the behavior of a local termination in the $q_1^*$ modulation. The beginning of the $q_1^*$ modulation is difficult to disentangle from the QPI due to the $q_1$ vector existing near/on top of $q_1^*$.

$<\Delta_1>$ is the average peak energy in the LDOS(E) and represents a maximum in the intensity of the pseudogap states[18]. If this LDOS(E) peak locally defines the pseudogap, there will be a peak in the $S_{x,y}$ intensity at the energy of the peak in the $\Delta_1$ distribution. In the second column of figure 14, both the $S_{x,y}$ intensity and the distribution of $\Delta_1$ are plotted for the three lowest doped samples. These plots show that at the energy where the maximum area of the field of view is at its LDOS(E) peak, there is a maximum in the intensity of $S_{x,y}$. This is also true if one compares $\Delta_1$ distribution with $Q_{x,y}$[12]. This implies that the $S_{x,y}$ signature of the pseudogap is linked locally to the peak in the LDOS(E) at the energy $\Delta_1$. In the past this has been assumed when the published Z-maps have had their $\Delta_1$ values normalized[7,12].

### 6b. $q_1^*$ modulation of the LDOS(*r*,E)

The energy scales $\Delta_0$ and $\Delta_{00}$ bracket the region where the $q_1^*$ modulation and the LDOS(E) kink exist. In the extremely underdoped spectra, the fitting implies that $\Delta_{00}$ and $\Delta_0$ are modulated in phase by the $q_1^*$ modulation (figure 6, *p* = 0.08, and *p* = 0.06). The $q_1^*$ modulation of these energy scales could be an actual modulation, an artifact caused by the $q_1^*$ modulating the local number of states in this energy range rather than the energy scales themselves, an effect of the setup condition, or a combination thereof. The measured kink size, $\Delta_{00} - \Delta_0$, shows a lack of the $q_1^*$ modulation signal (figure 6). This implies that the $q_1^*$ modulation is likely not an intrinsic modulation in $\Delta_0$ and $\Delta_{00}$, unless there is a well-defined modulation in the superconducting energy scale that is in phase and identical in magnitude to a $\Delta_{00}$ modulation.

To determine between the possible scenarios an adaptation of the lock-in technique[48] is used to create an amplitude normalized $q_1^*$ real space pattern from the UD20 K data (the spatial lock in parameters used are shown in supplementary figure 12). The $q_1^*$ modulation pattern is taken at an energy of 40 meV, which is midway between the average value of $<\Delta_0>$ = 23 meV and below the ending average, $<\Delta_{00}>$ = 60 meV. Using the amplitude normalized $q_1^*$ map, the average of the spectra is calculated as a function of the $q_1^*$ normalized amplitude. In figure 15a an example of the normalized $q_1^*$ modulation is shown. The red section represents half the $q_1^*$ wavelength and the averaged LDOS(E) is calculated over this region, from the trough of the $q_1^*$ modulation to the crest. This average LDOS(E) is shown in figure 15b, where there is a clear uniform increase in the kink height from the trough of the $q_1^*$ pattern to the crest. The Tripartite model fitting detects this rise in the kink as a shift of $\Delta_0$ in sync with a $\Delta_{00}$ shift. If one accepts that the low energy (below $\Delta_0$) coherent momentum Eigen states of Bogoliubov quasiparticles are uniform in overall number of states[4], then this analysis shows the effects of the setup condition on the measured LDOS (E), since in figure 15b below $<\Delta_0>$ all the LDOS(E) curves should be coincident. In order to correct for this setup condition effect, while retaining positive and negative data, one could set the number of states at $\Delta_0$ to be the same across the field of view.

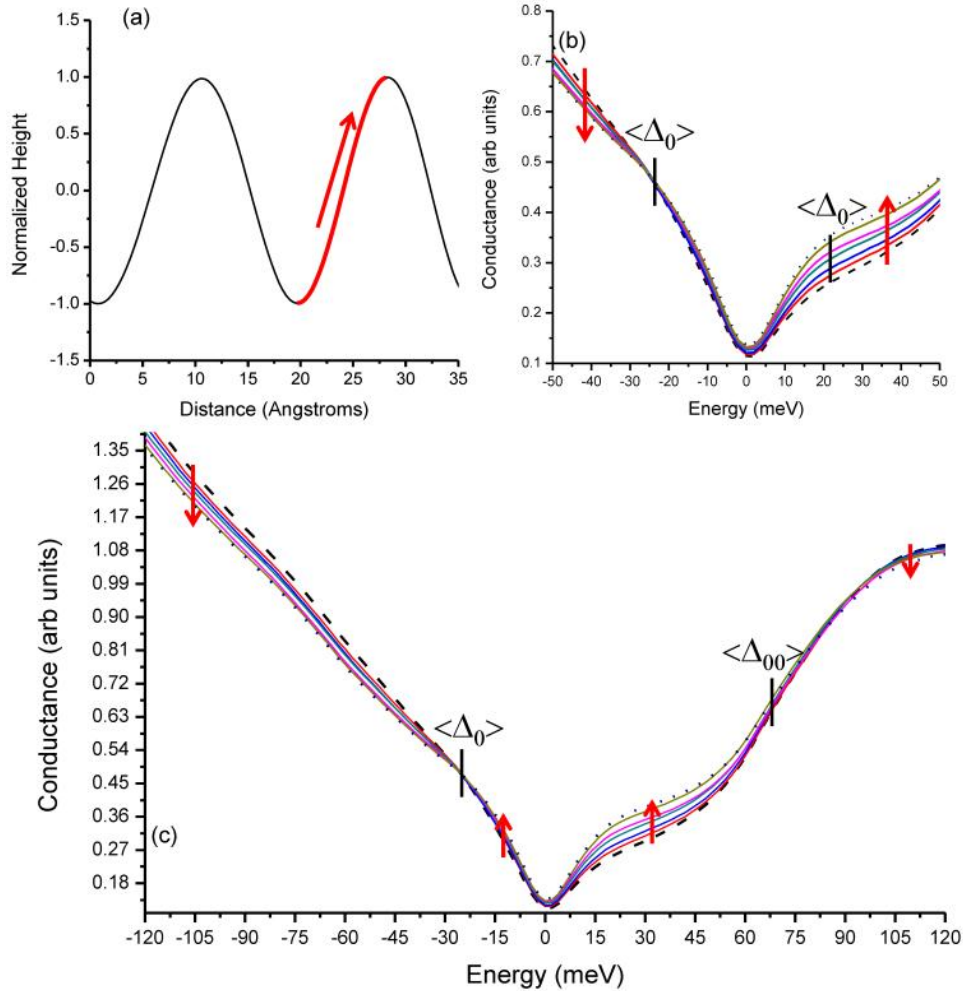

**Figure 15:** The $q_1^*$ modulations effect on the measured LDOS(E). (a) Shows a single slice of the $q_1^*$ modulation extracted from UD20 K sample. This was accomplished by a similar method to the supermodulation studies reported previously[48]. Averaging the LDOS(*r*,E) over the red section, or one half of the oscillation, allows the effect of the $q_1^*$ modulation on the LDOS(E) to be plotted in (b). The red arrows show the progression of the LDOS(E) curves going from trough to crest. (c) Plots the same average for a larger energy range, showing the transitions that happen for all three regions and the effect the Z-map would have on the $q_1^*$ modulations energy range and intensity. This figure demonstrates how the fitting of $\Delta_0$ and $\Delta_{00}$ will pick up the $q_1^*$ modulation, and also how the Z-map will remove/suppress the $q_1^*$ modulation everywhere but between $\Delta_0$ and $\Delta_{00}$. This implies that the $q_1^*$ modulation is constrained to the kink area and everywhere else it is seen is due to a setup effect. Due to limited data, it is unknown if the $q_1^*$ modulation exists at negative energies as well. The lock in techniques parameters can be seen in supplementary figure 12.

SI-STM works by changing the distance between the SI-STM tip and sample until it reaches a constant current at the voltage where the dI/dV sweep begins. For small energies this is represented by

$$\frac{dI(\bm{r},E)}{dV} = \frac{eI_s N(\bm{r},E)}{\int_0^{eV_s} N(\bm{r},\omega)d\omega} \qquad 8$$

where $N(\bm{r},E)$ is the density of states, $V_s$ is the setup voltage, $I_s$ is the setup current and e is the charge of an electron. If there is a modulation in the total amount of states from point to point that have energies between zero and the setup voltage, the STM will decrease or increase the tip sample separation in order to maintain constant setup current. This is because the setup current at a set tip sample spacing is proportional to the number of states between zero and the setup voltage. If there is a modulation in the total number of states from location to location, then the STM will modulate the distance in order to maintain a constant current. Since the setup current is only on one side of the spectrum, in this analysis the positive side, the change in the setup condition will have a similar effect on the opposite side of the spectrum where those extra states may not exist, but will now show up in the data due to this modulation. A phase modulation in the LDOS(E) that exists on both sides of zero bias is a possible setup condition effect. When the two modulations are out of phase there exists an actual LDOS(E) modulation. By taking the ratio of the positive energies to the negative energies, these 'setup effect' states with the same phase should be canceled out[7] and the LDOS(E) modulations will remain. Taking the ratio of Eq 8 one can see that the integrated LDOS(E) will cancel out. Of course if there is an LDOS(E) modulation that exists in phase on both sides of zero bias, then it will be canceled out in the Z-map as well. This is why $q_5$ in the dispersive QPI is suppressed when the ratio map is taken[5,11,49].

In Figure 15b,c at energies below $<\Delta_0>$ the positive and negative LDOS(E) both increase from the trough to crest of the $q_1^*$ modulation. Taking the ratio of the positive and negative energies will cancel out the majority of this modulation, which is why the Z-map is used to remove the $q_1^*$ modulation for QPI studies[5]. Between $<\Delta_0>$ and $<\Delta_{00}>$, the LDOS(E) modulation has an opposite phase for the positive and negative energies, implying that there is a LDOS(E) modulation in the number of states on the positive side of the spectrum due to the $q_1^*$ modulation between these two energies. This is consistent with the $q_1^*$ modulation being a positive energy phenomenon in the Fourier transform of the LDOS($r$,E)[11]. Above $\langle\Delta_{00}\rangle$ the modulation order is reversed on the positive side of the spectrum, at these energies the Z-map will cancel out a large part of the $q_1^*$ modulation. This shows that the presence of the $q_1^*$ modulation at $\Delta_1$ is likely a setup effect. The only energy range where the positive and negative energies are out of phase is between $\Delta_0$ and $\Delta_{00}$, proving that the modulation in the number of states happens between these energy values. The reason why, in figure 15, the number of states caused by the setup effect appears not to completely cancel out in the low energy and high energy regions is likely due to the disorder in $\Delta_{00}$, $\Delta_0$, and $\Delta_1$ which are not taken into account in this analysis. Also there is likely an effect of the dispersive $q_1^*$ vector whose presence will cause some interference when it is near $q_1^*$ in wavelength[11].

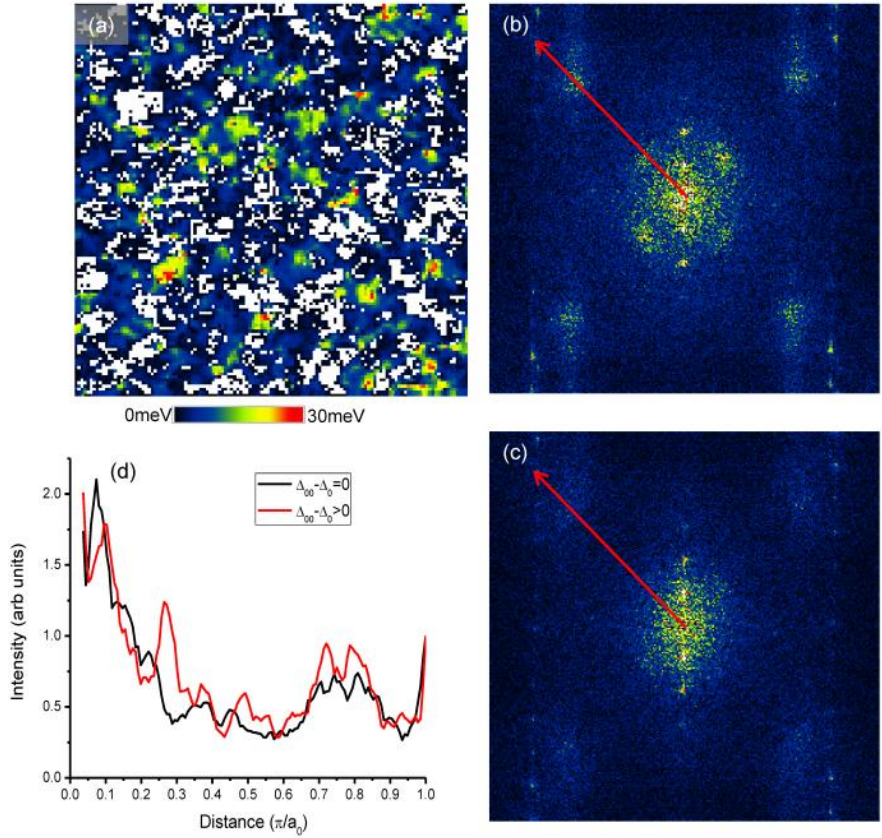

**Figure 16:** The $q_1^*$ modulation exists only where the kink exists. (a) $\Delta_{00} - \Delta_0$ map for UD74 K data with white set to locations where $\Delta_{00} - \Delta_0 = 0.0$. The data set is separated into two regions and Fourier transformed. In the region where the kink exists, (b), there is a clear $q_1^*$ pattern in the raw data. Where the kink does not exist, (c), shows no $q_1^*$ peak in the raw data. The FT was taken at 38 meV. Cuts of the intensity along the atomic direction (red arrow) in (d) display the removal of the $q_1^*$ peak in the region without the kink.

There is a kink present in the filled state LDOS(E) (negative energies) as well[3], at moderately underdoped samples it appears as a plateau, but at the low dopings, in figure 15c, it is represented as inflection point in the large background slope. There is a lack of another change over in the order of the LDOS(E) at the negative energies $\langle\Delta_{00}\rangle$. This could be due to setup effects overwhelming a small signal there, the lack of the pseudogap states, or a negative energy increase in $\Delta_{00}$ beyond the energy range of the data. There is also a lack of a $q_1^*$ vector at negative energies[11], but this could be due to the data set being set up at positive energy bias. In order to confirm the lack of a filled state $q_1^*$ modulation as well investigate $\Delta_{00}$ and $\Delta_1$, data must be taken at negative energy setup voltages and compared to positive setup voltages. This would also require a negative energy model. Both this data and model are currently not available.

Using the data and the Tripartite model parameterization, another feature of the checkerboard that has been reported on in the past[3] can be clarified. Previously it was shown that by masking the data into high $\Delta_1$ and low $\Delta_1$ regions, a field of view can be segregated into a region with and without the checkerboards signal. In this paper we have shown that the $q_1^*$ modulation exists between $\Delta_0$ and $\Delta_{00}$. Since $\Delta_{00}$'s energy dependence tracks $\Delta_1$ and as $\Delta_1$ increases $\Delta_{00} - \Delta_0$ increases, the relatively large $\Delta_1$ sections of the field of view should contain the $q_1^*$ modulation. The previous study uses this dependence to separate the $q_1^*$ signal areas. In figure 16a the same analysis is carried out, however instead of using $\Delta_1$, the kink/$q_1^*$ modulations energy span $\Delta_{00} - \Delta_0$ is used. Applying the same masking technique, the field of view is separated into two regions, $\Delta_{00} - \Delta_0 > 0$ and $\Delta_{00} - \Delta_0 = 0$. If the kink is the LDOS(E) marker of the $q_1^*$ modulation, then the $q_1^*$ signature will only be present in the $\Delta_{00} - \Delta_0 > 0$ section of the data. In figure 16 a sample where the doping permits large regions with and without the kink (UD74 K, $p = 0.14$) is divided into the two regions. The Z-map is used in order to remove the setup effect of the $q_1^*$ modulation from the FT. Figure 16(b-c) confirms that the $q_1^*$ peak exists only in the region $\Delta_{00} - \Delta_0 > 0$. This provides another conformation that the kink is the LDOS(E) signature of the $q_1^*$ modulation. A further $q_1^*$ modulation example using the $\Delta_1$ dependence of $\Delta_0$ and $\Delta_{00}$ is presented in supplementary figure 13.

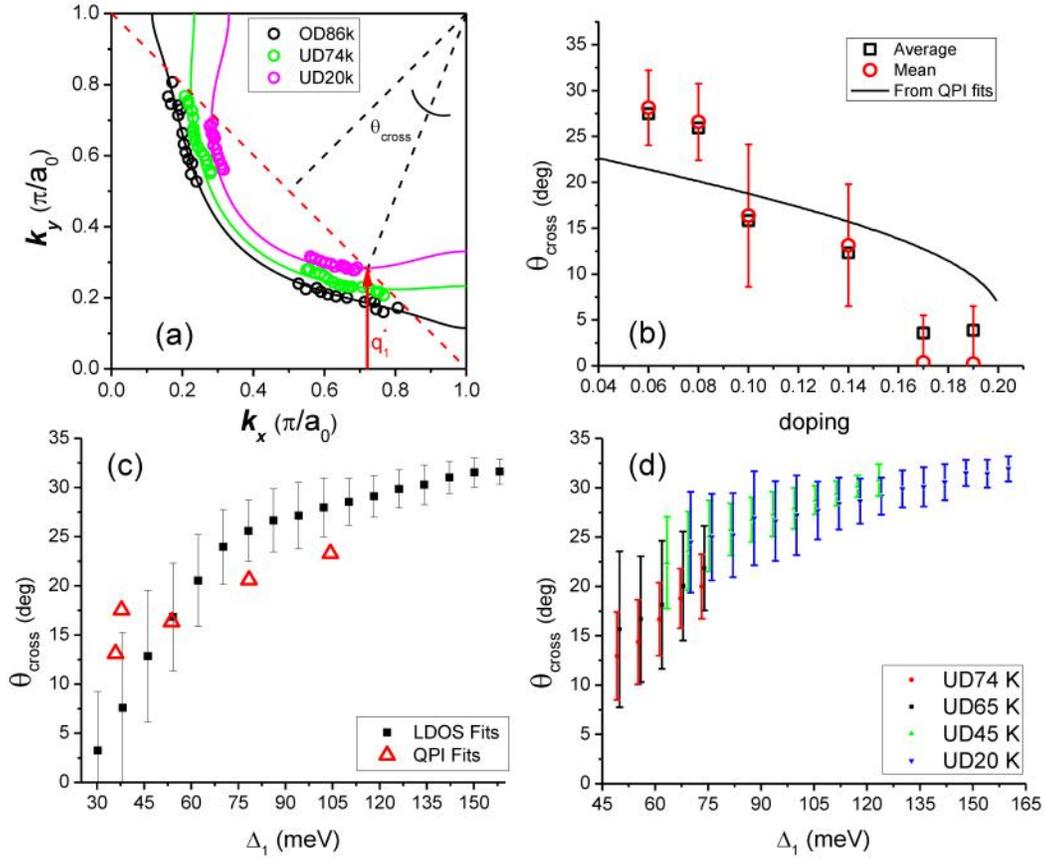

**Figure 17:** Dependence of $\theta_{cross}$ on doping showing that the local $\theta_{cross}$ has an effective local band structure dependence. a) QPI extracted $k$-space origins for three of the dopings and tight binding band structure fits. The dashed red line shows the PAF-zone boundary, and $\theta_{cross}$ is illustrated for the UD20 K data. The red arrow shows the direction and size of half the $q_1^*$ $q$-vector for UD20 K and as the doping is increased the wave vector will decrease following the intersection of the Fermi surface with the PAF-zone boundary. b) plots $\theta_{cross}$ for each of the six samples, error bars are the standard deviation for each doping. The sold line is the expected $\theta_{cross}$ for the PAF-zone boundary from a rigid band structure fit of the dispersive QPI. c) is the dependence of $\theta_{cross}$ on $\Delta_1$ for all the data. The average is taken excluding those regions where $\theta_{cross} = 0.0$. The error bars are the standard deviation for each point. The open red triangles represent the $\theta_{cross}$ at the last point where a QPI signal can be measured. d) is the same graph as in b) but differentiated between the four lowest dopings showing the continuous doping change.

## 6c. Parent Compounds AF-Zone boundary

We have shown that the three energy scales dependence on the doping inferred from $\Delta_1$ is singularly valued. This relationship is numerically the same as that previously[11] calculated using only the mean LDOS(E) and the dispersive QPI (see figure 13b). In the mean LDOS(E) and dispersive QPI measurements, the kink starts at the dispersive QPI termination energy and ends at the -3 dB point of the $q_1^*$ modulation. The $q_1^*$ modulation begins and the dispersive QPI termination occur at the location in $k$-space where the Fermi surface reaches the PAF-zone boundary. This is illustrated in Figure 17a, where the dispersive QPI extracted k-space origin points, with their tight binding band structure fits are plotted. The last point measured is just before the PAF-zone boundary, marked in red. The wavelength of the $q_1^*$ and $S_{x,y}$ modulations are also set by this crossing point[10,12]. Figure 17a shows half the $q_1^*$ vector in red for the UD20 K data and as the doping increases the vector shrinks in size as it follows the intersection of the PAF-zone boundary and the Fermi surface. In this local study, there is a disorder in the energy at which the kink is observed. This leads to the question of how the PAF-zone boundary is related to the local disorder in the kinks energy location.

There are two likely possibilities for the relationship between the spatially disordered kink and the PAF-zone boundary. The first scenario is that the kinks energy location is consistent with the dispersive QPI measured Fermi surface PAF-zone termination. That is the kinks energy is set by a given samples local $\Delta_1$ value and that samples Fermi surface. In this case $\theta_{cross}$ should be a constant at each doping and not vary from patch to patch. The second scenario is that the patchwork represents a distribution across different effective local dopings and the kinks position is set by not only the local $\Delta_1$ value, but by that patches equivalent band structure for a singly doped, homogeneous sample. In this case $\theta_{cross}$ should vary from patch to patch within each sample, while being continuous across all samples. Each patch with the same $\Delta_1$ value will have the same $\theta_{cross}$ and its value should scale with local effective doping.

The 2nd scenario is supported by previous work that shows the $q_1^*$ modulations wavelength varies locally and that its wavelength depends on the local $\Delta_1$ value of a patch in Bi-2201[50]. Although here we have shown that the $q_1^*$ modulation is confined to a specific energy range. This variation in the $q_1^*$ modulations wavelength was attributed to local changes in the Fermi surface. Figure 17a shows how this scenario would work, if $\theta_{cross}$ changes locally, then the intersection with the PAF-zone boundary will change, and the resulting $q_1^*$ wave vector will change accordingly. The 2nd scenario is also supported by the continuous variation in $\Delta_0$, $\Delta_{00}$, with $\Delta_1$ between different samples. If the kinks energy location was set by the dispersive QPI derived Fermi surface and $\Delta_1$ then there should be jumps in the dependence of $\Delta_0$ and $\Delta_{00}$ on $\Delta_1$ between differently doped samples.

Figure 17(b-d) plots the $\theta_{cross}$ as a function of doping for six differently doped samples as well $\theta_{cross}$ as a function of $\Delta_1$. Figure 17b shows that at high dopings and low dopings the expected $\theta_{cross}$ from the QPI derived band structure is undershot and overshot by the local fits kink locations. For the high dopings this could be due to the LDOS(E) $\Delta_1$ peak overlapping the kink and preventing its accurate measure. This supports the theory that the superconducting and pseudogap energy scales do not merge and instead overlap[18,51,52] near $p \sim 0.19$. For the extremely underdoped samples the kink lies at a larger angle than expected. This could be due to the kink starting further away from the PAF-zone boundary due to an increase in anti-ferromagnetic correlations at low dopings[53,54].

The individual dopings (figure 17b) and the continuous $\Delta_1$ dependence of $\theta_{cross}$ (Figure 17c,d) follow the same trend as seen by the dispersive QPI data. However the kinks position changes continuously and single valued along the same curve with decreasing doping/increasing $\Delta_1$. $\theta_{cross}$ as a function of $\Delta_1$ is calculated by the ratio of $\Delta_{00}/\Delta_1$ in the Tripartite model (see eq. 7). Overall $\theta_{cross}$ is not dependent on the sample doping/Fermi surface, but changes throughout each sample in a manner consistent with a continuously changing PAF-zone boundary intersection driven by Fermi surface that varies from patch to patch. This agrees with the measurement of the local variance in q1* measured in Bi-2201[50]. This continuous change suggests that each patch can be modeled as having the equivalent electronic structure as a homogeneous sample consisting of a single $\Delta_1$ value and doping. The underlying parameter that determines each patches electronic structure would then be an effective doping. This effective doping would set both $\Delta_1$ and the band structure parameters which determine $\theta_{cross}$.

## 7. Conclusion

We have demonstrated the ability to effectively parameterize the LDOS(***r***,E) data with the Tripartite model not only globally, as previously reported, but locally with atomic resolution. This allows the detailed study of the local variation of the multiple energy scales and their associated ***q***-space phenomena for the first time. The data analyzed allows for a more complete picture of the SI-STM data to be reported. This study leaves the higher energy boson mode[25] and the negative energy, filled state, LDOS(E) for future research. The main obstacle to both projects is the lack of a model that can be fitted to those data sets accurately. Here, we have also demonstrated the ability to fit a computationally intense model to large SI-STM datasets. This allows for future studies involving theories/models that have been previously considered too intensive for localized fitting.

The local disorder in the three energy scales seen in $Bi_2Sr_2CaCu_2O_{8+x}$ describes a patchwork superconductor, where each individual patch has the LDOS(E) of an equivalent sample that has a single doping and is homogeneous. This picture of the patchwork and the universal nature of the disorder require that there be an underlying scattering or disorder mechanism that is independent of doping or dopant atoms. This mechanism causes a local change in the LDOS(E) equivalent to a change in the local effective hole number. One possible candidate for the origin of this disorder is cation disorder which is present in this family of cuprates[55]. Regardless of the specific cause of the disorder, the mechanism by which this disorder affects the low energy states can take the form of a change in the local a pairing interaction[36,42], or a superexchange variation[23,56].

The $\Delta_0$ and $\Delta_{00}$ maps have the same spatial disorder as the $\Delta_1$ map and there exists a continuous and singly valued relationship between their values and $\Delta_1$. We also find that all three energy scales can be defined by a single underlying parameter that has a doping independent spatial disorder, where the bulk doping value only serves to shift the average of this underlying parameter. The kinks energy position and span change continuously with increasing $\Delta_1$, implying that the kink phenomena and the linked $q_1^*$ modulation excitation are not tied to the Fermi surface measured from the dispersive QPI, but to a local effective Fermi surface for each patch. The underlying parameter is consistent with a local effective doping level that uniquely defines a Fermi surface, $\Delta_1$, $\Delta_{00}$, and $\Delta_0$ for each patch. However this model only works by modeling the local patches as an infinite homogeneous superconductor of only that local doping. It is also possible the some underlying scatterer is acting to change the phase of the wave function and variations in the strength of the scatterer are causing variations in this phase shift. The wave length of the interference pattern we measure as $q_1^*$ would then change depending on local changes in this phase shift.

The Tripartite models application to LDOS(***r***,E) data sets allows the determination of the relationship between the spatial excitations and the patchwork variation of energy scales. The QPI termination is driven by the ending of the superconducting state[57] and the $q_1^*$ modulation is resolved as a positive energy phenomenon that results in the modulation of the number of states in the LDOS(E)'s kink. The $\Delta_{00}$ distribution shows that the $q_1^*$ modulation terminates locally as each patch passes its $\Delta_{00}$ energy, making the $q_1^*$ modulations termination a local phenomenon. The pseudogap peak at $\Delta_1$ is correlated with the intensity of $S_{x,y}$ and $Q_{x,y}$ linking the peak in the LDOS(E) to the short wavelength pseudogap modulation and marking the pseudogap a phenomena local to each patch as well.

The exact nature of the $q_1^*$ modulation is still unknown and there are many theories about the pseudogap state and its associated smectic stripes. The nature of the state between the superconducting and the pseudogap state, with a separate excitation that is linked to the PAF-zone boundary and a locally determined Fermi surface, is still a mystery. This state could be driven by AF fluctuations[58,59] or scattering between the ends of the band structure[60], however the local change observed here and elsewhere[17] of the $q_1^*$ modulations effective band structure is unexplained. The determination that the $q_1^*$ modulation is linked locally to the kink, as well as the determination of its energy span with doping will aid in the determination of origin and identification of this state.

The lifetime broadening of the $\Delta_1$ peak is set not only by the energy of $\Delta_1$, but also by the boundaries between the different patches. These boundaries do not change with doping, and the lifetime increase at them must either be caused by the interaction between the patches, or by an underlying mechanism that defines the patchwork disorder. This could be due to differences in the local pairing strength driven by disorder in the pairing mechanism[23], or disorder outside the CuO plane[45] effecting the states that lie in plane through a distortion of the unit cell[61]. The lifetime change with doping inside the patches is caused either the establishment of the nematic and smectic pseudogap states or a change in the dopant disorder/screening length[13] once the doping has dropped below a certain threshold.

By measuring the three energy scales present in the sample as a function of doping inferred from the $\Delta_1$ value, we can provide a diagram of the electronic structure as a function of doping. In figure 18, we show in red the coherent Bogoliubov quasiparticles excitations of the superconducting state, representing the dispersive QPI and the V shaped LDOS(E). These states are terminated collectively across all the patches and are replaced by individual patchwork excitations of the $q_1^*$ modulation/kink shown in green. The $q_1^*$ modulation/kink is bounded by $\Delta_0$ and $\Delta_{00}$ and the energy range over which it is found grows larger with decreasing doping. Above the $q_1^*$ modulation, the pseudogap states exist with their peak in energy at $\Delta_1$ shown here in blue. The pseudogap states peak locally with the LDOS(E) peak, however like the $q_1^*$ modulation, their excitation pattern of $S_{x,y}$ and $Q_{x,y}$, is spatially localized across the field of view[8,9]. This doping driven diagram shows the evolution of the E < 150 meV positive LDOS(E) for underdoped $Bi_2Sr_2CaCu_2O_{8+x}$ and our Tripartie model represents a method of reducing the LDOS(E) to its key observables and showing that there is one underlying parameter, with a doping independent spatial disorder, that determines the underdoped electronic states.

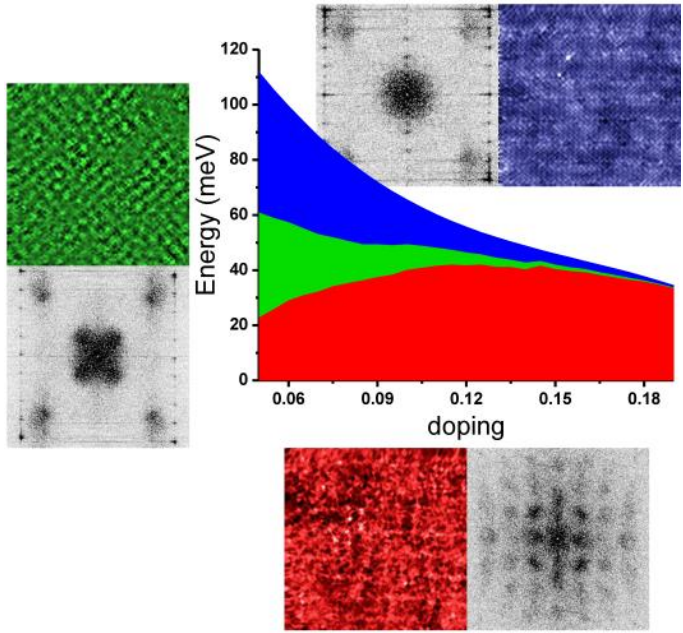

**Figure 18:** Doping dependence of all three phases extracted from the data. At low energies, corresponding to the superconducting dome, the coherent dispersive Bogoliubov quasiparticles shown in red exist. These states terminate at $\Delta_0$. The real space and the *q*-space patterns are shown below the phase diagram (real space on left, *q*-space on right). The $q_1^*$ modulation regime bridges the energy range between the pseudogap and the superconductivity energy scales. The $q_1^*$ modulations real space and *q*-space patterns are shown to the left and are marked in green; here the low wavelength ($q < 0.10/a_0$) frequencies have been removed for clarity. Above the $q_1^*$ modulation in energy, the pseudogap states with real space and *q*-space patterns shown above the phase diagram and is marked in blue.

**Acknowledgements:** The authors wish to acknowledge the support of J.C. Davis in allowing us to use the data and in manuscript proofreading help, as well as discussions. We would like to acknowledge the support of John Moreland. We would also like to thank Inês Firmo for proofreading help and we would like to acknowledge Magus for insightful conversations.

We would like to thank the NVIDIA Academic Hardware program for providing hardware for this project and the Sloane Foundation for providing funding.

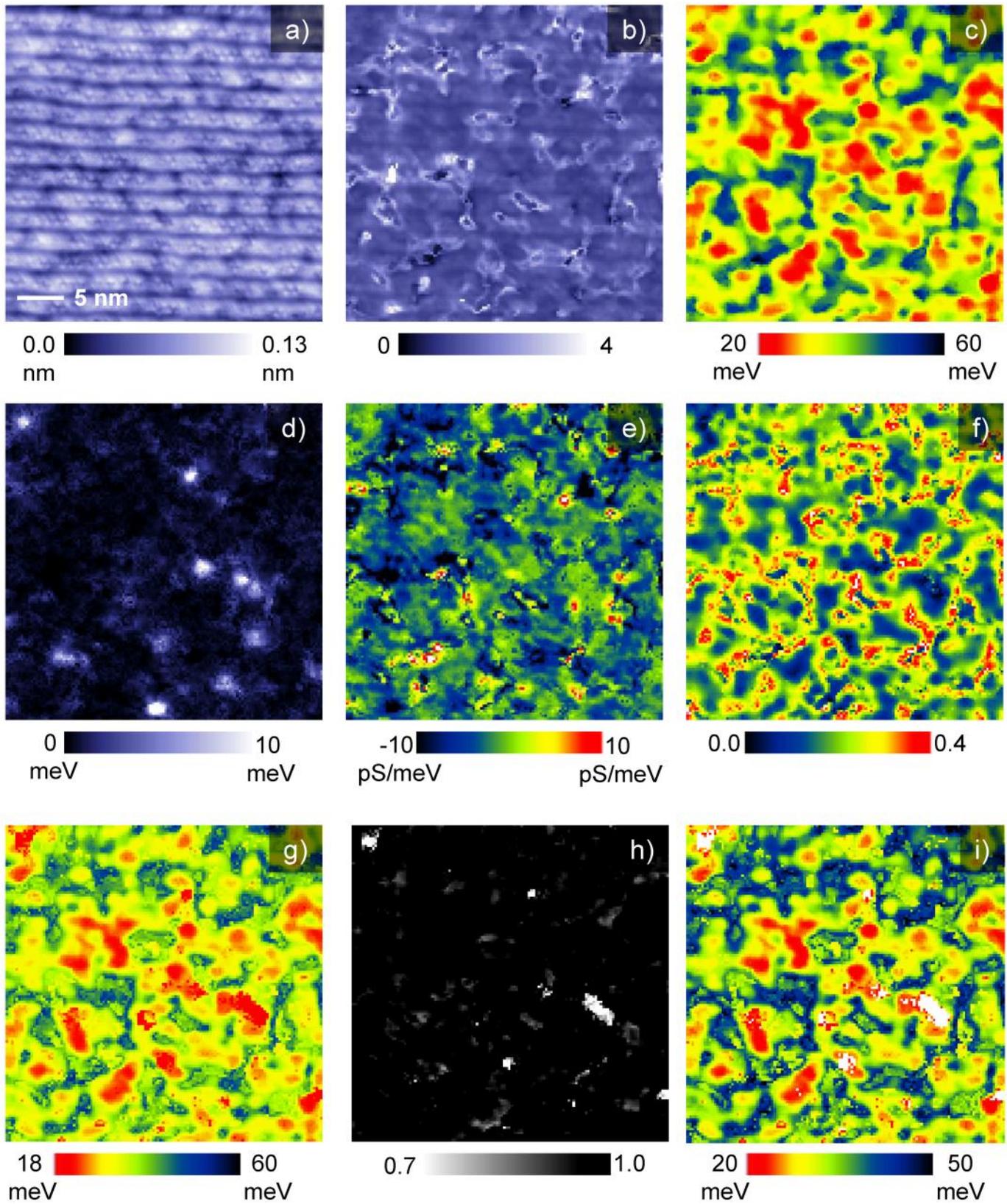

**Supplementary Figure 1:** OD86 K complete fit and topograh. The OD86 K data set is a 40.2 nm² field of view. a) shows the simultaneous topograph taken with the spectroscopic data. b) is the prefactor for the fits, c) the $\Delta_1$ map, d) $\Gamma_1$ map, e) slope, f) $\alpha$ map, g) $\Delta_{00}$ map, h) B map, and i) $\Delta_0$ map.

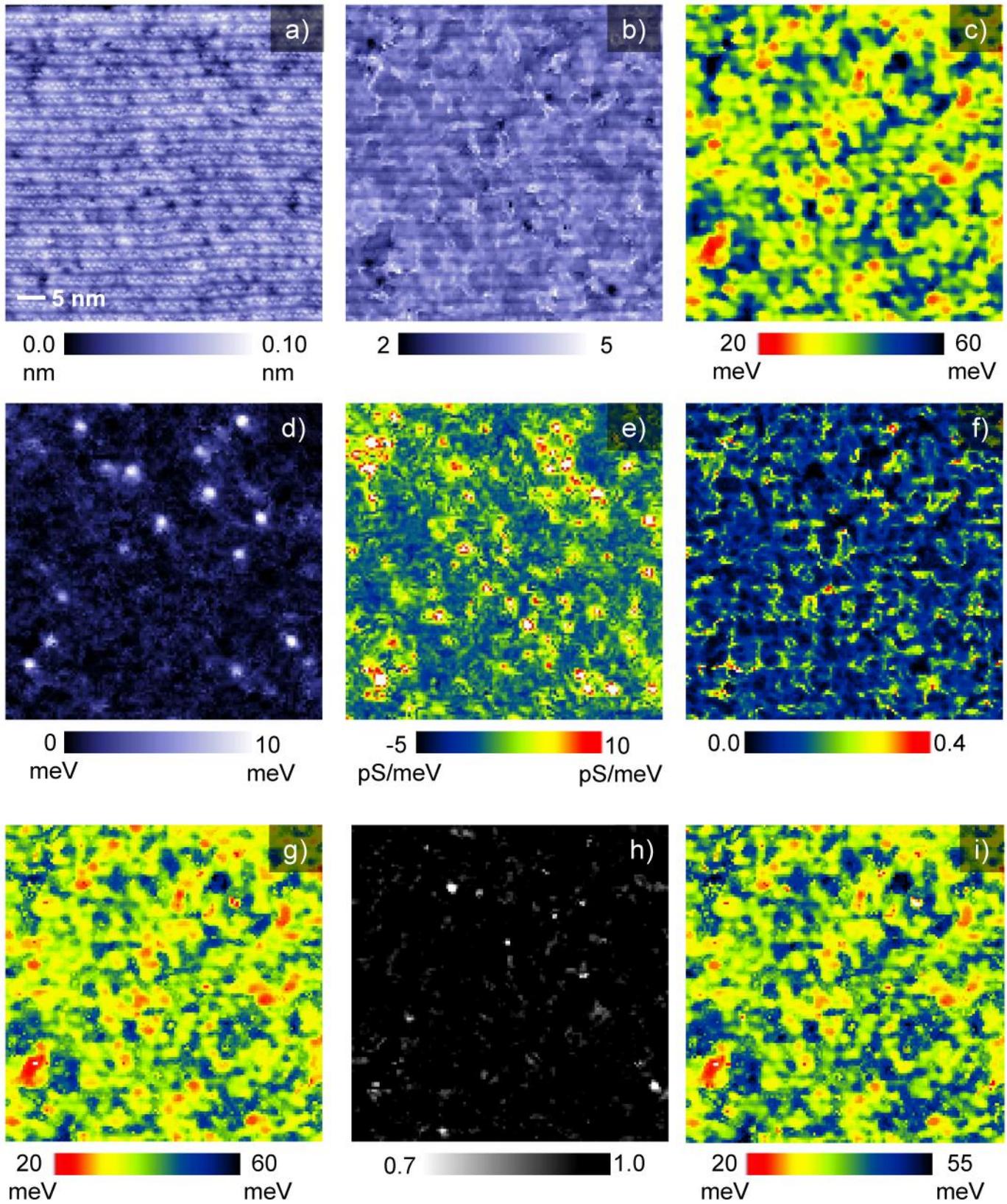

**Supplementary Figure 2:** OD89 K complete fit and topograph. The OD89 K data set is a 55 nm$^2$ field of view. a) shows the simultaneous topograph taken with the spectroscopic data. b) is the prefactor for the fits, c) the $\Delta_1$ map, d) $\Gamma_1$ map, e) slope, f) $\alpha$ map, g) $\Delta_{00}$ map, h) B map, and i) $\Delta_0$ map.

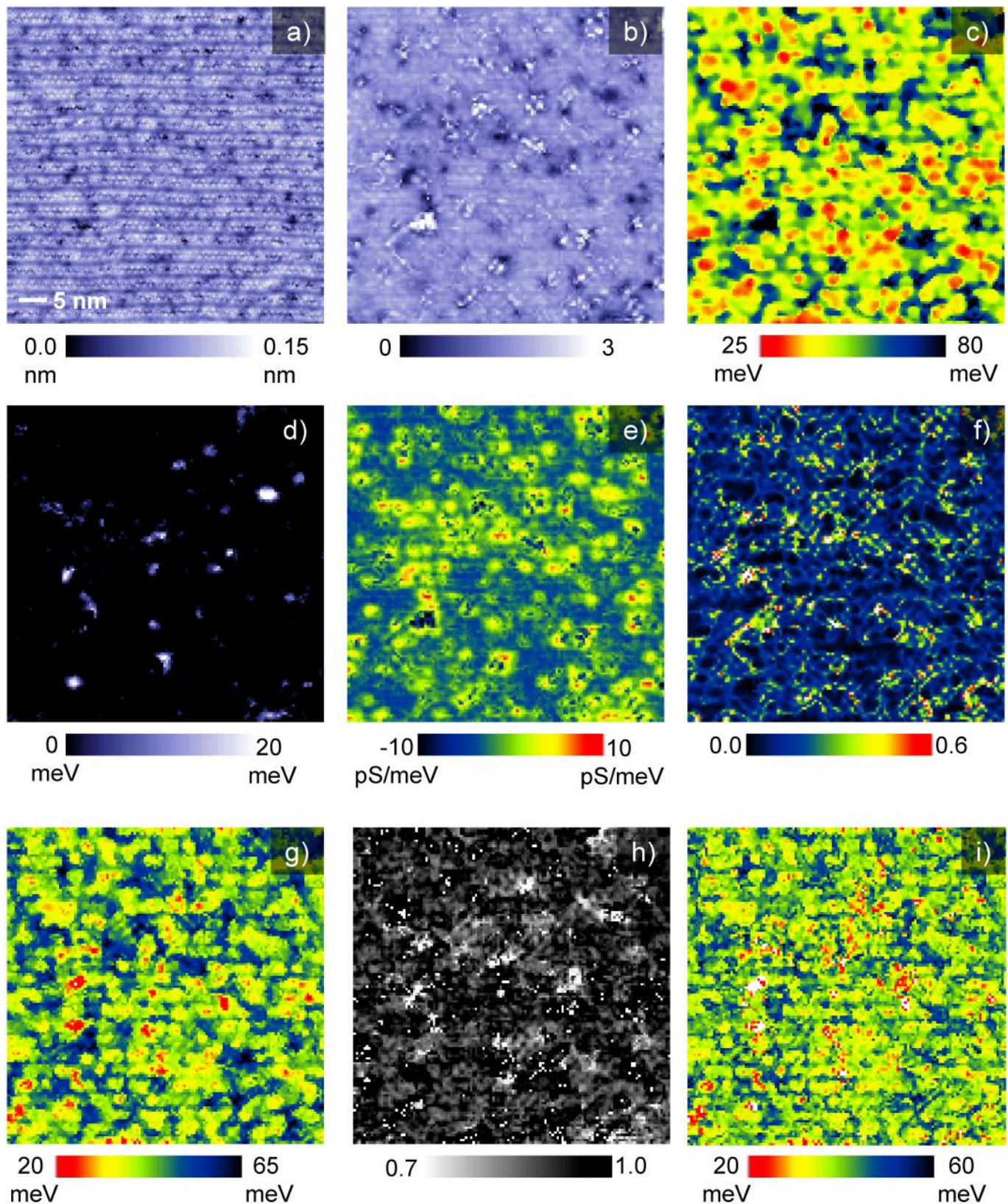

**Supplementary Figure 3:** UD74 K complete fit and topograph. The UD74 K data set is a 55 nm² field of view. a) shows the simultaneous topograph taken with the spectroscopic data. b) is the prefactor for the fits, c) the $\Delta_1$ map, d) $\Gamma_1$ map, e) slope, f) $\alpha$ map, g) $\Delta_{00}$ map, h) B map, and i) $\Delta_0$ map.

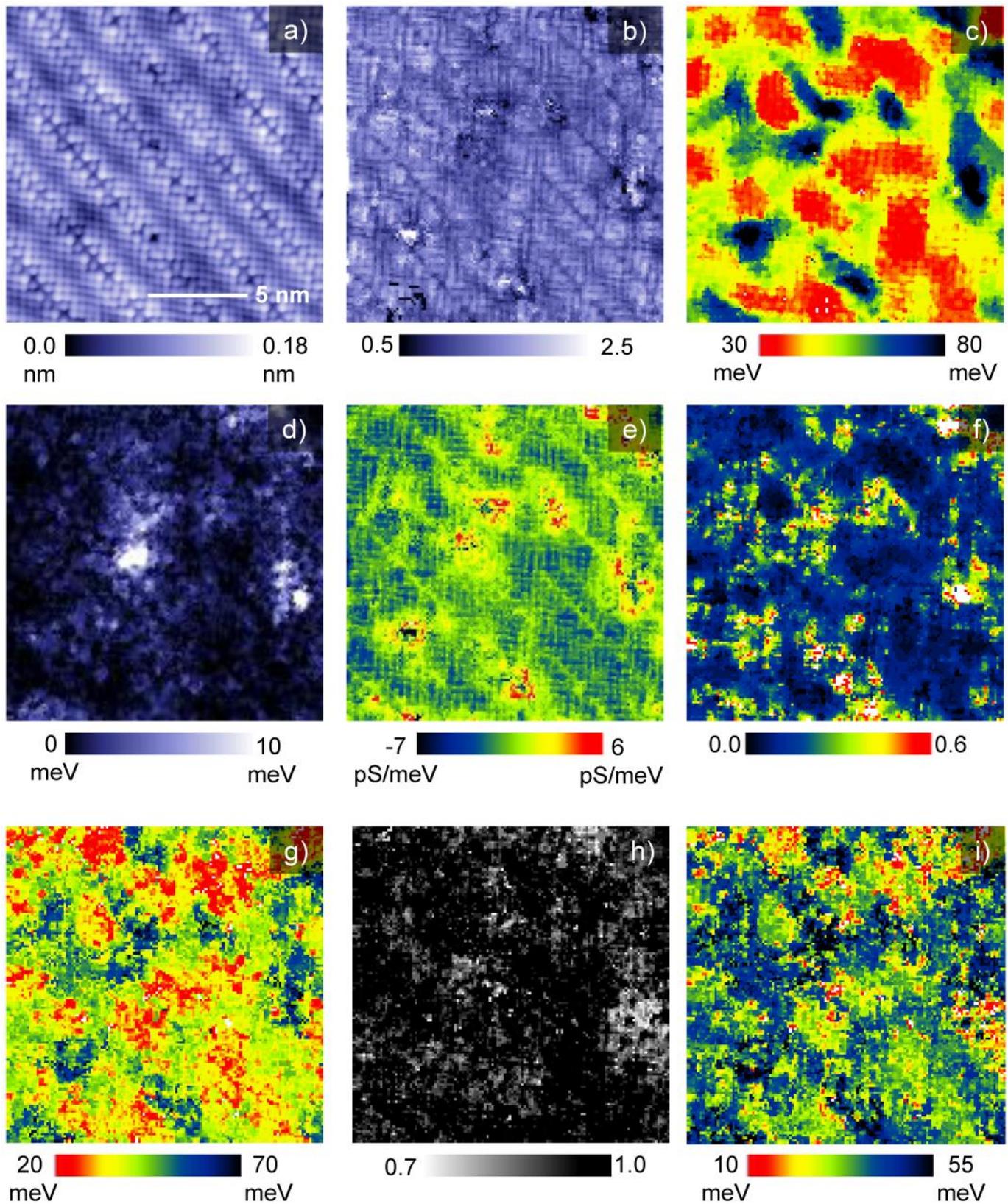

**Supplementary Figure 4:** UD74 K small area complete fit and topograph. The UD74 K data set is a 15.4 nm² field of view. a) shows the simultaneous topograph taken with the spectroscopic data. b) is the prefactor for the fits, c) the $\Delta_1$ map, d) $\Gamma_1$ map, e) slope, f) $\alpha$ map, g) $\Delta_{00}$ map, h) B map, and i) $\Delta_0$ map.

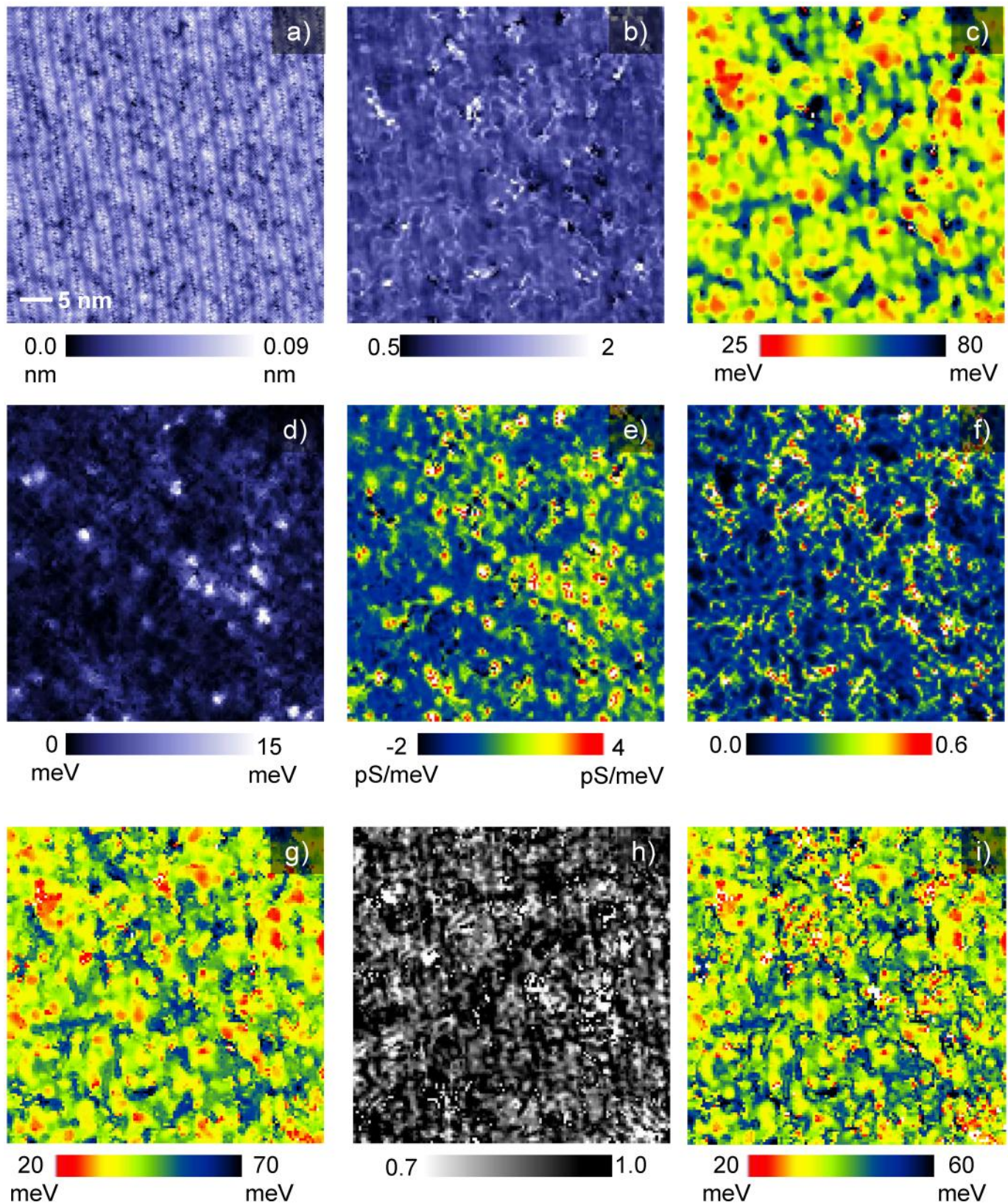

**Supplementary Figure 6:** Ni-UD74 K oxygen complete fit and topograph. The UD74 K data set is a 48 nm² field of view. The nickel impurities cause increased error in the fits due to positive 10-40 meV resonances and there are ~40 nickel atoms in the field of view. a) shows the simultaneous topograph taken with the spectroscopic data. b) is the prefactor for the fits, c) the $\Delta_1$ map, d) $\Gamma_1$ map, e) slope, f) $\alpha$ map, g) $\Delta_{00}$ map, h) B map, and i) $\Delta_0$ map.

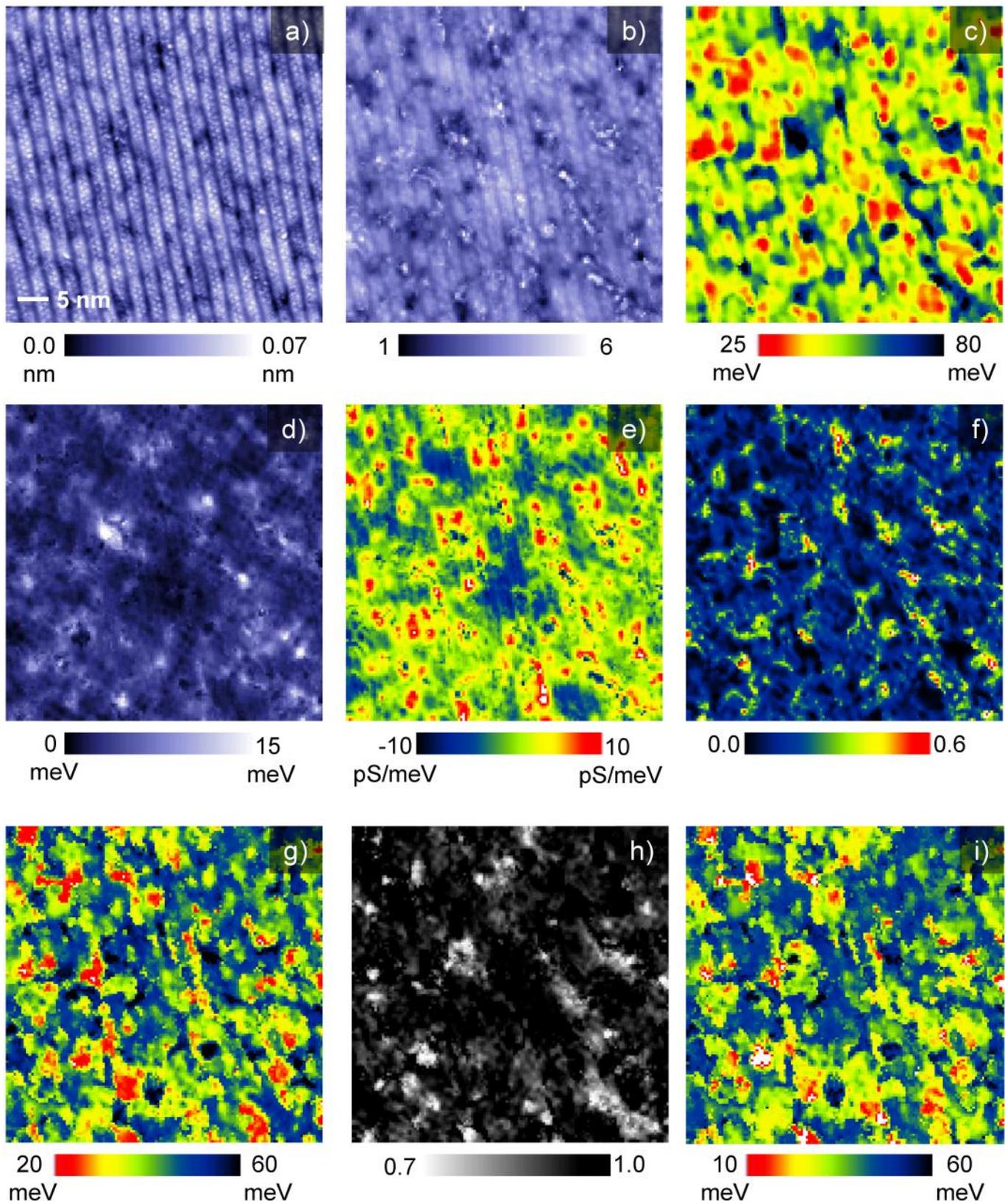

**Supplementary Figure 7:** UD65 K complete fit and topograph. The UD65 K data set is a 55 nm$^2$ field of view. a) shows the simultaneous topograph taken with the spectroscopic data. b) is the prefactor for the fits, c) the $\Delta_1$ map, d) $\Gamma_1$ map, e) slope, f) $\alpha$ map, g) $\Delta_{00}$ map, h) B map, and i) $\Delta_0$ map.

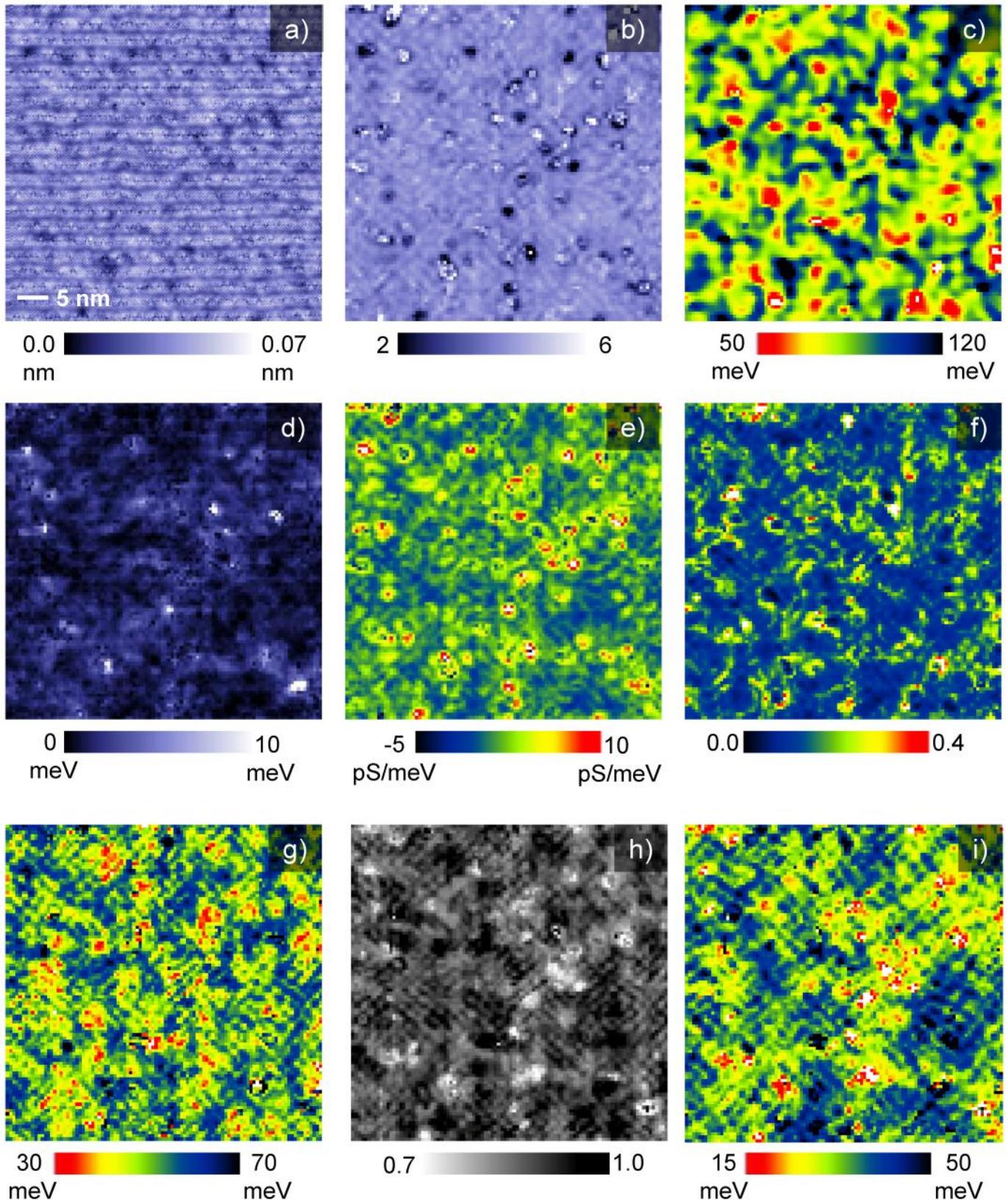

**Supplementary Figure 8:** UD45 K complete fit and topograph. The UD45 K data set is a 50 nm² field of view. a) shows the simultaneous topograph taken with the spectroscopic data. b) is the prefactor for the fits, c) the $\Delta_1$ map, d) $\Gamma_1$ map, e) slope, f) $\alpha$ map, g) $\Delta_{00}$ map, h) B map, and i) $\Delta_0$ map.

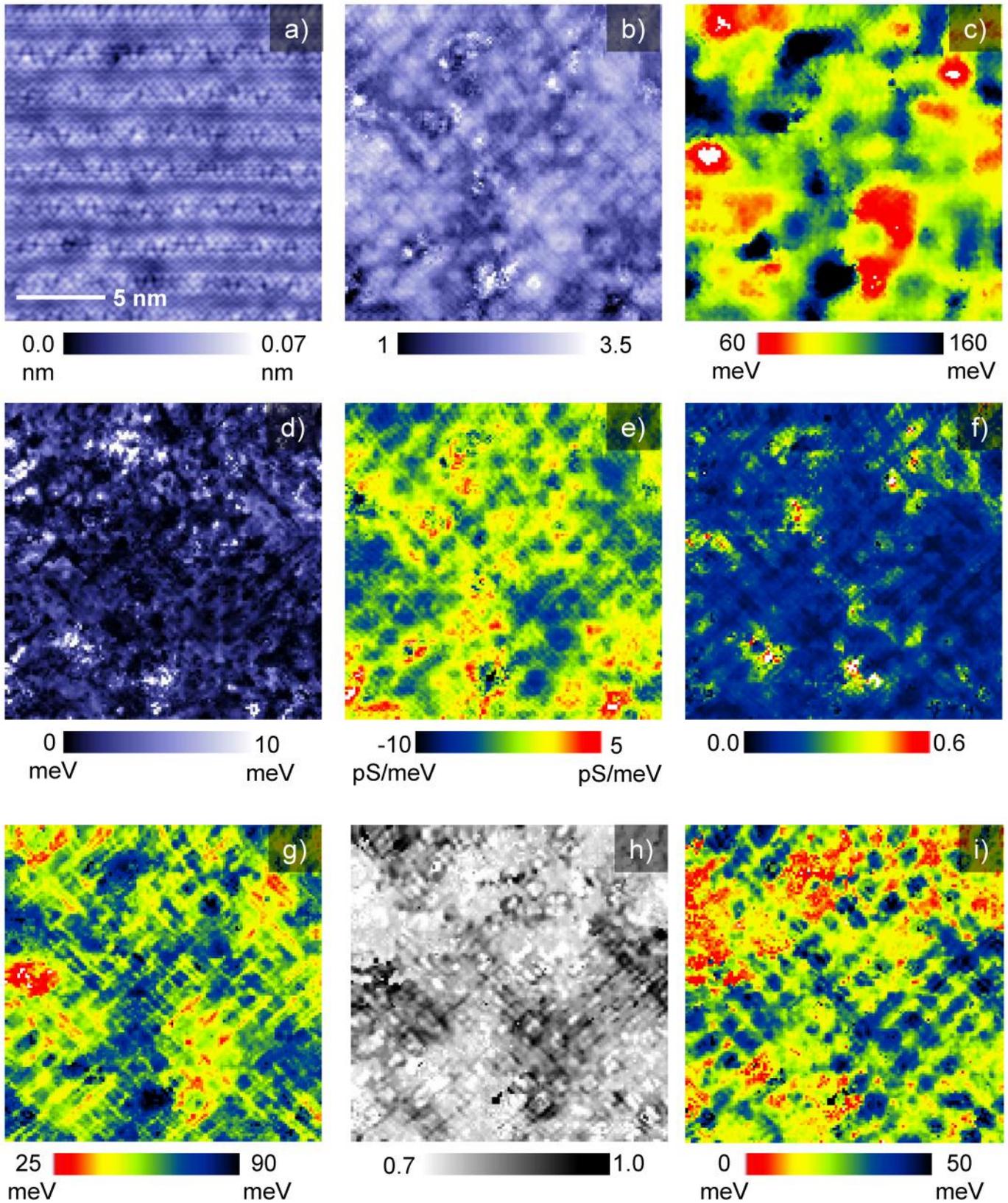

**Supplementary Figure 9:** UD20 K complete fit and topograph. The UD20 K data set is an 18 nm² field of view. a) shows the simultaneous topograph taken with the spectroscopic data. b) is the prefactor for the fits, c) the $\Delta_1$ map, d) $\Gamma_1$ map, e) slope, f) $\alpha$ map, g) $\Delta_{00}$ map, h) B map, and i) $\Delta_0$ map.

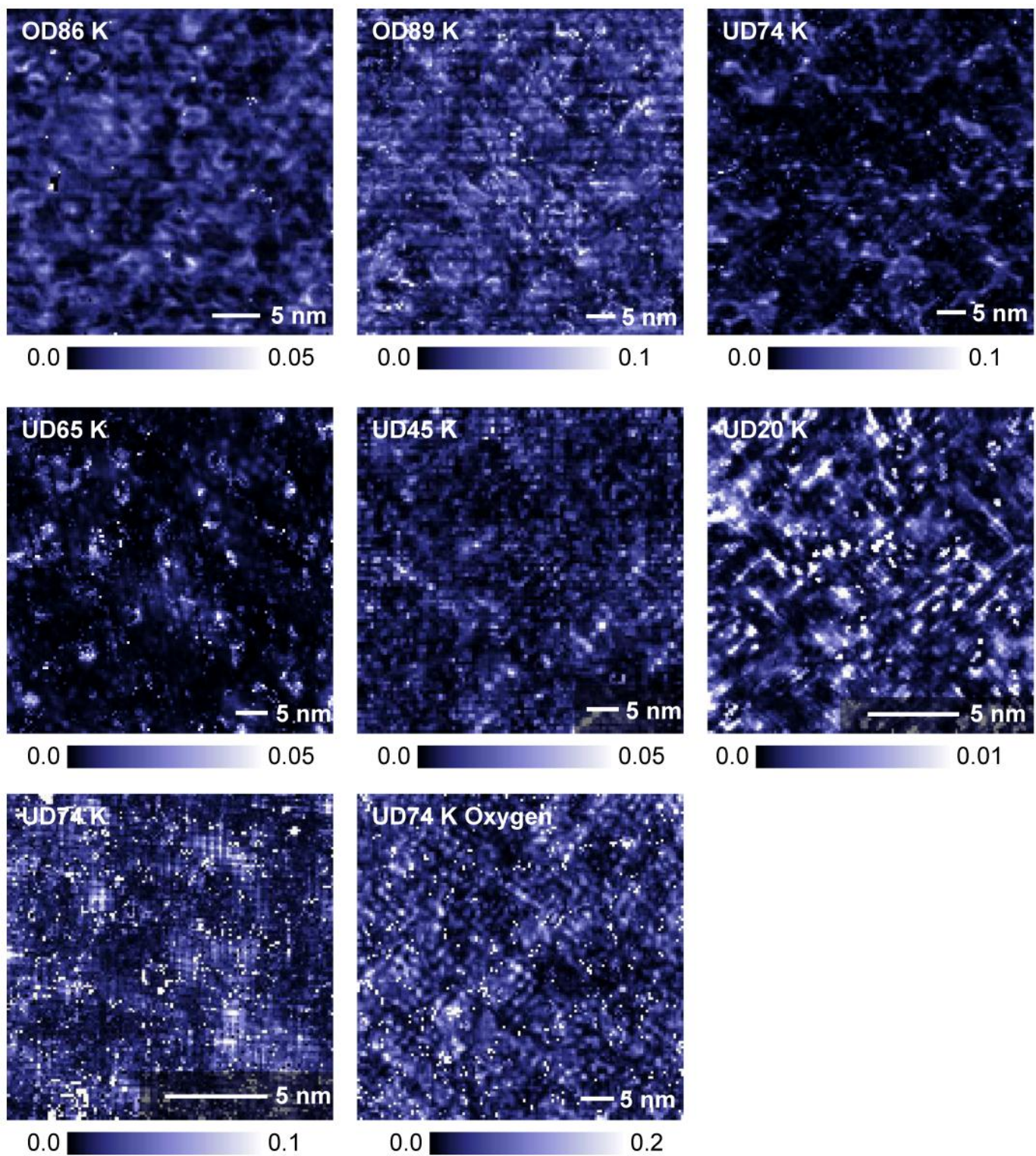

**Supplementary Figure 10:** Normalized $\chi^2$ for all fits.

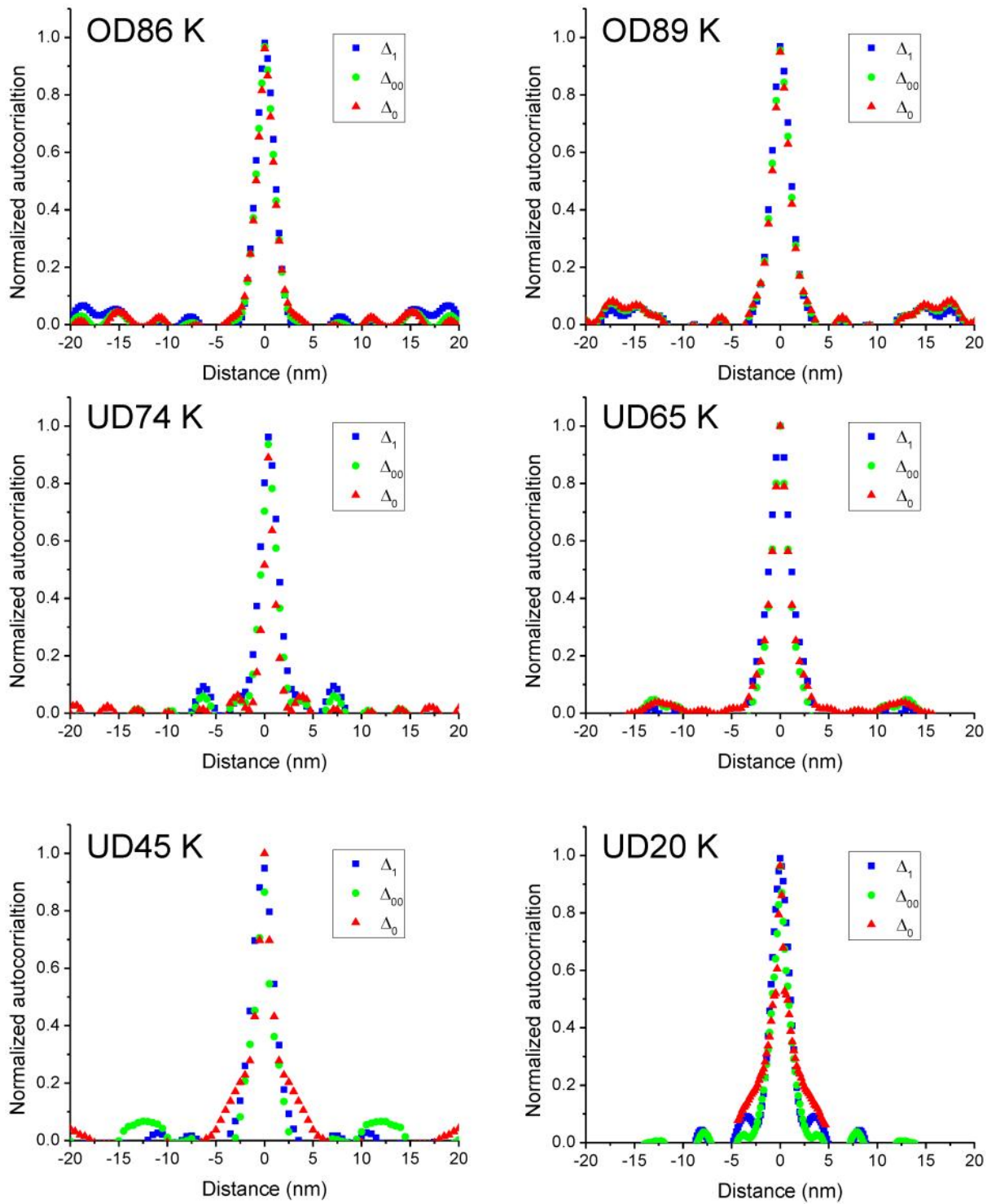

**Supplementary figure 11:** Normalized autocorrelations of all three energy scales for all three data sets. Due to the varying resolution of the original files, and the low resolution used in this study, the majority of autocorrelations are limited by resolution as far as noise. For UD45 K and US20 K the checkerboard is also a problem and it has been filtered out of the UD20 K data in an attempt to compensate. Also any pixels that exists as spikes due to noise and bad fits tends to accentuate the x = 0 nm peak resulting in bad fits for the FWHM displayed in table 2.

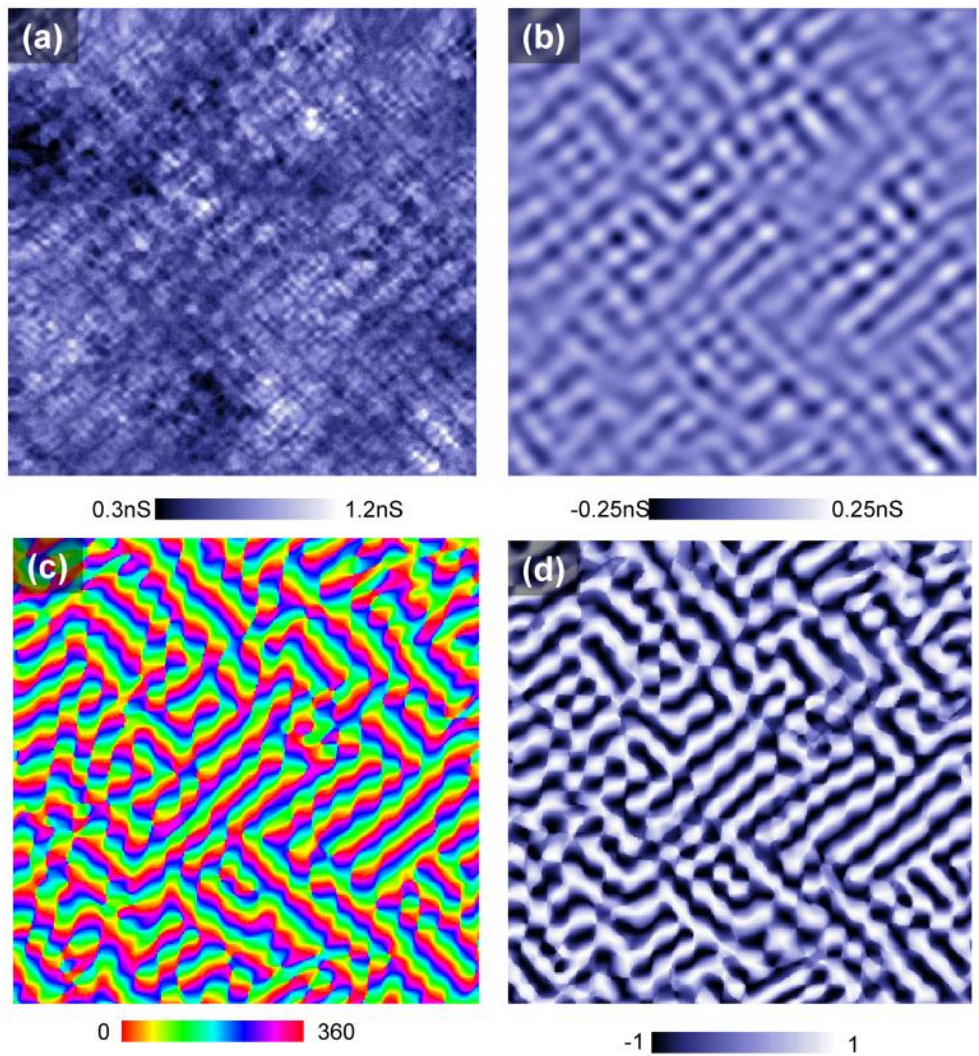

**Supplementary figure 13.** a) 40 meV layer for UD20 K. b) the checkerboard Fourier filtered. C) theta defined as the average of the two different direction theta. D) the normalized amplitude that can be used to generate figure 12.

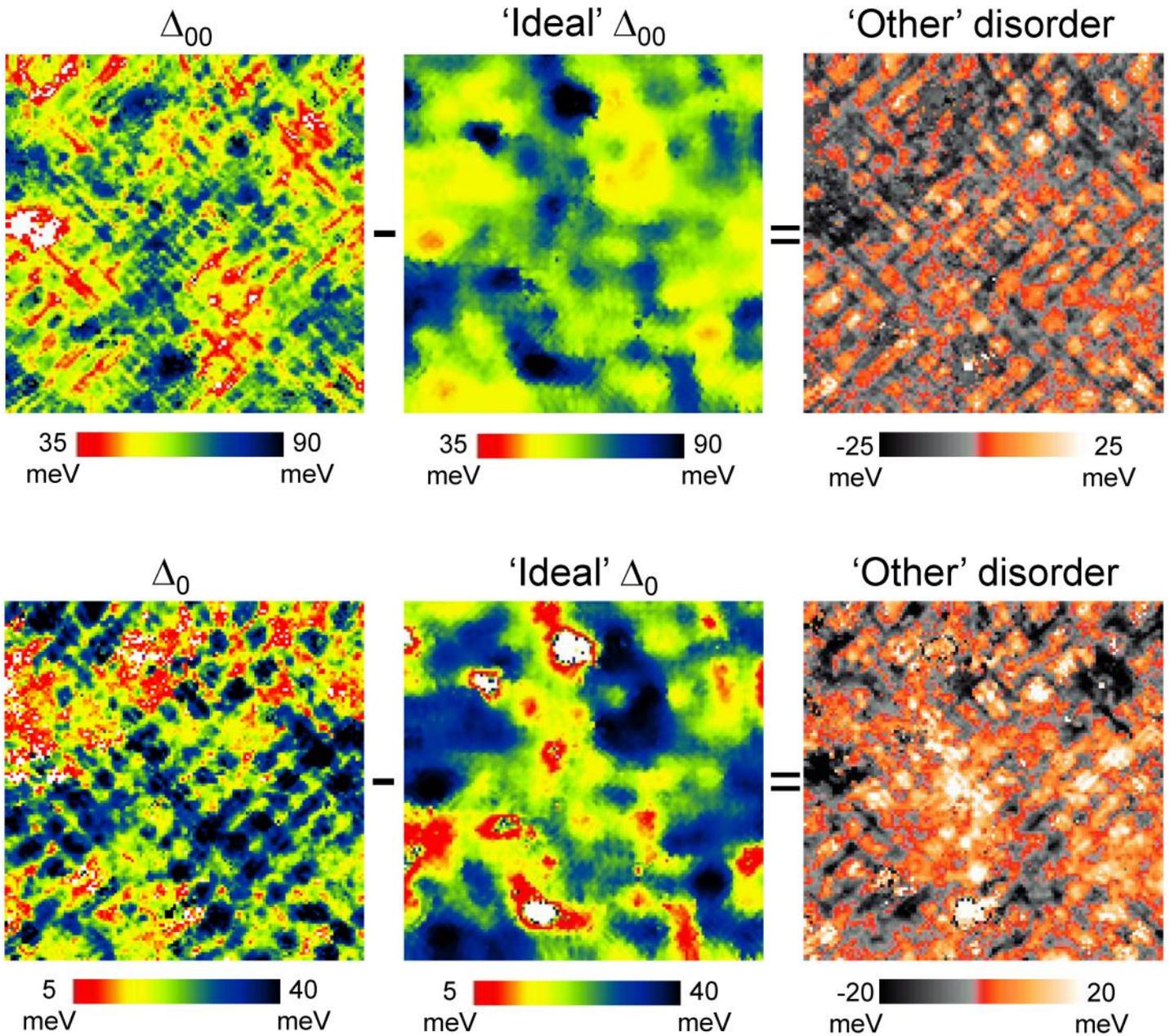

**Supplementary figure 13:** Checkerboard as the 'other disorder'. Using the relationships between (figure 8) $\Delta_1$, $\Delta_0$, $\Delta_{00}$ an 'ideal' $\Delta_0$ and $\Delta_{00}$ maps for a given $\Delta_1$ map are generated. These maps are 'ideal' in the sense of a single value for of $\Delta_0$ and $\Delta_{00}$ as determined by the local $\Delta_1$. If we apply this to the UD20 K data, we can see that the deviations from this 'ideal' model take the form of the checkerboard. The UD20 K was chosen due to the large kink and the accuracy to which we can fit the kink. The UD20 K data set also has a small field of view which helps isolate the exact signature of the 'other disorder'.